\documentclass[aps,prx,twocolumn,amsmath,amssymb,showpacs,superscriptaddress,notitlepage,longbibliography]{revtex4-1}
\usepackage[colorlinks=true,linkcolor=blue,anchorcolor=red,citecolor=blue, urlcolor=blue]{hyperref}
\usepackage{bm}
\usepackage{graphicx}
\usepackage{color}
\usepackage{subfigure}
\usepackage{amsmath, bm}
\usepackage{makecell}
\usepackage{graphicx,xcolor,framed}
\usepackage{amsmath}
\colorlet{shadecolor}{gray!10}
\newcommand{\BOX}[1]{\vskip 2mm\begin{shaded}\it #1\end{shaded}\vskip -2mm}

\begin{document}

\title{Perspective: Nonlinear Hall Effects}

\author{Z. Z. Du}
\affiliation{Shenzhen Institute for Quantum Science and Engineering and Department of Physics, Southern University of Science and Technology (SUSTech), Shenzhen 518055, China}
\affiliation{Shenzhen Key Laboratory of Quantum Science and Engineering, Shenzhen 518055, China}

\author{Hai-Zhou Lu}
\email{Corresponding author: luhz@sustech.edu.cn}
\affiliation{Shenzhen Institute for Quantum Science and Engineering and Department of Physics, Southern University of Science and Technology (SUSTech), Shenzhen 518055, China}
\affiliation{Shenzhen Key Laboratory of Quantum Science and Engineering, Shenzhen 518055, China}

\author{X. C. Xie}
\affiliation{International Center for Quantum Materials, School of Physics, Peking University, Beijing 100871, China}
\affiliation{CAS Center for Excellence in Topological Quantum Computation, University of Chinese Academy of Sciences, Beijing 100190, China}
\affiliation{Beijing Academy of Quantum Information Sciences, West Building 3, No.10, Xibeiwang East Road, Haidian District, Beijing 100193, China}

\date{\today }

\begin{abstract}
The Hall effects comprise one of the oldest but most vital fields in condensed matter physics, and they persistently inspire new findings such as quantum Hall effects and topological phases of matter.
The recently discovered nonlinear Hall effect is a new member of the family of Hall effects. It is characterised as a transverse Hall voltage in response to two longitudinal currents in the Hall measurement, but it does not require time-reversal symmetry to be broken. It has deep connections to symmetry and topology and thus opens new avenues via which to probe the spectral, symmetry and topological properties of emergent quantum materials and phases of matter. In this Perspective, we present an overview of the recent progress regarding the nonlinear Hall effect. After a review of the theories and recent experimental results, we discuss the problems that remain, the prospects of the use of the nonlinear Hall effect in spectroscopic and device applications, and generalisations to other nonlinear transport effects.
\end{abstract}

\maketitle

The Hall effects refer to a transverse voltage in response to longitudinal currents, and they represent important paradigms in condensed matter physics \cite{Hall1879,Hall1881,Klitzing80prl,Tsui82prl,Chang13sci}. Most Hall effects result from broken time-reversal symmetry, which is induced by magnetic fields \cite{Prange12book} or magnetic orders \cite{Nagaosa10rmp,Hall1881,Nakatsuji2015Nature,Machida2010Nature,Yasuda16np}, as constrained by Onsager's reciprocal theorem \cite{Onsager1931PR1}, see Box.~1.
There have been mainly two ways to realise a Hall effect without  breaking time-reversal symmetry: generalising to spin \cite{Sinova15rmp} or valley \cite{Xiao2007PRL,Yao2008PRB,Xiao2012PRL,Mak2014Sci} degrees of freedom  or transcending the linear-response regime.

\begin{figure*}[htpb]
\centering
\includegraphics[width=0.98\textwidth]{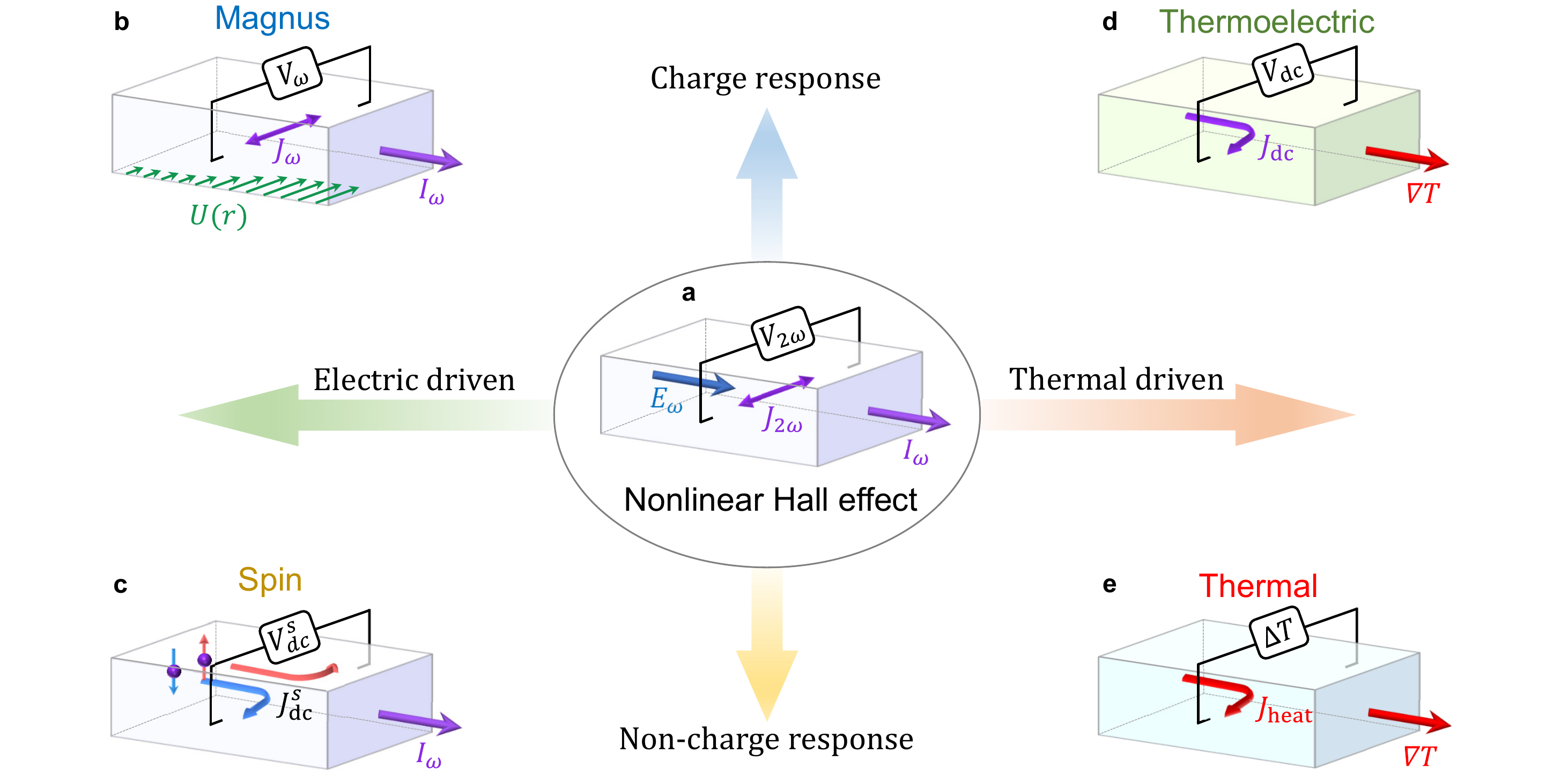}
  \caption{\textbf{The nonlinear Hall effect and its generalisations.} \textbf{a}, Schematic of how to measure the nonlinear Hall effect. Using the lock-in technique, a transverse voltage $V_{2\omega}$ is measured in response to an ultralow-frequency $ac$ currents $I_{\omega}$. Theoretically, the voltage is formulated quadratically as a transverse current density $J_{2\omega}$ in response to the electric field $E_\omega$ induced by $I_{\omega}$. $2\omega$ can be generalised to $\omega_1\pm \omega_2$, that is, input currents with different frequencies. One can achieve the gyrotropic Hall effect \cite{Konig19prb} if one of the frequencies is zero.
  The generalization of the nonlinear Hall effect can be achieved by replacing the electric driving field or charge response current, as shown in (\textbf{b}-\textbf{e}) for
several examples of generalization.
  \textbf{b}, In the Magnus Hall effect,  one of the driving fields is replaced by the gradient of an electrostatic potential $U(\bm{r})$ \cite{Papaj19prl,Mandal2020PRB}.
  \textbf{c}, In the nonlinear spin Hall effect, a nonlinear spin Hall current is driven in response to the electric driving field \cite{Hamamoto2017PRB,Araki2018SR}.
  By replacing the driving electric field with a temperature gradient $\nabla T$, one may have the nonlinear thermoelectric (\textbf{d}) and nonlinear thermal Hall (\textbf{e}) effects \cite{Nakai2019PRB,Su19prb,Zeng19prb,Zeng2020PRR}, for the electric and thermal Hall response currents, respectively.}\label{Fig:nonlinear Hall effect}
\end{figure*}

The recently discovered nonlinear Hall effect
\cite{Sodemann15prl,Low15prb,Facio18prl,You18prb,ZhangY18-2dm,Zhang18PRB,Du18prl,Ma19nat,Kang19nm}
is characterised by a transverse voltage in response to two longitudinal driving currents.
The nonlinear Hall effect does not require breaking time-reversal symmetry but inversion symmetry. This presents new avenues via which to probe the spectral, symmetry and topological properties of emergent quantum phases of matter and materials, and this effect can be generalised to more unconventional responses when breaking discrete and crystal symmetries, or to related phenomena, such as the nonlinear spin Hall effect \cite{Hamamoto2017PRB,Araki2018SR}, the gyrotropic Hall effect \cite{Konig19prb}, the Magnus Hall effect \cite{Papaj19prl} and the nonlinear Nernst effect \cite{Su19prb,Zeng19prb} (Fig.~\ref{Fig:nonlinear Hall effect}), thus leading to a fertile ground for future research .

In this Perspective, our aim is to present the recent progress in the field of nonlinear Hall effects. We begin with a brief introduction of the measurement of the nonlinear Hall effect,
followed by the geometric and disorder-induced contributions to the effect, the scaling law of the effect and symmetry analysis. As an outlook, we review avenues for future research, including spectroscopic and device applications and generalisations to related nonlinear Hall phenomena.

\begin{figure*}[htpb]
\centering
  \includegraphics[width=1\textwidth]{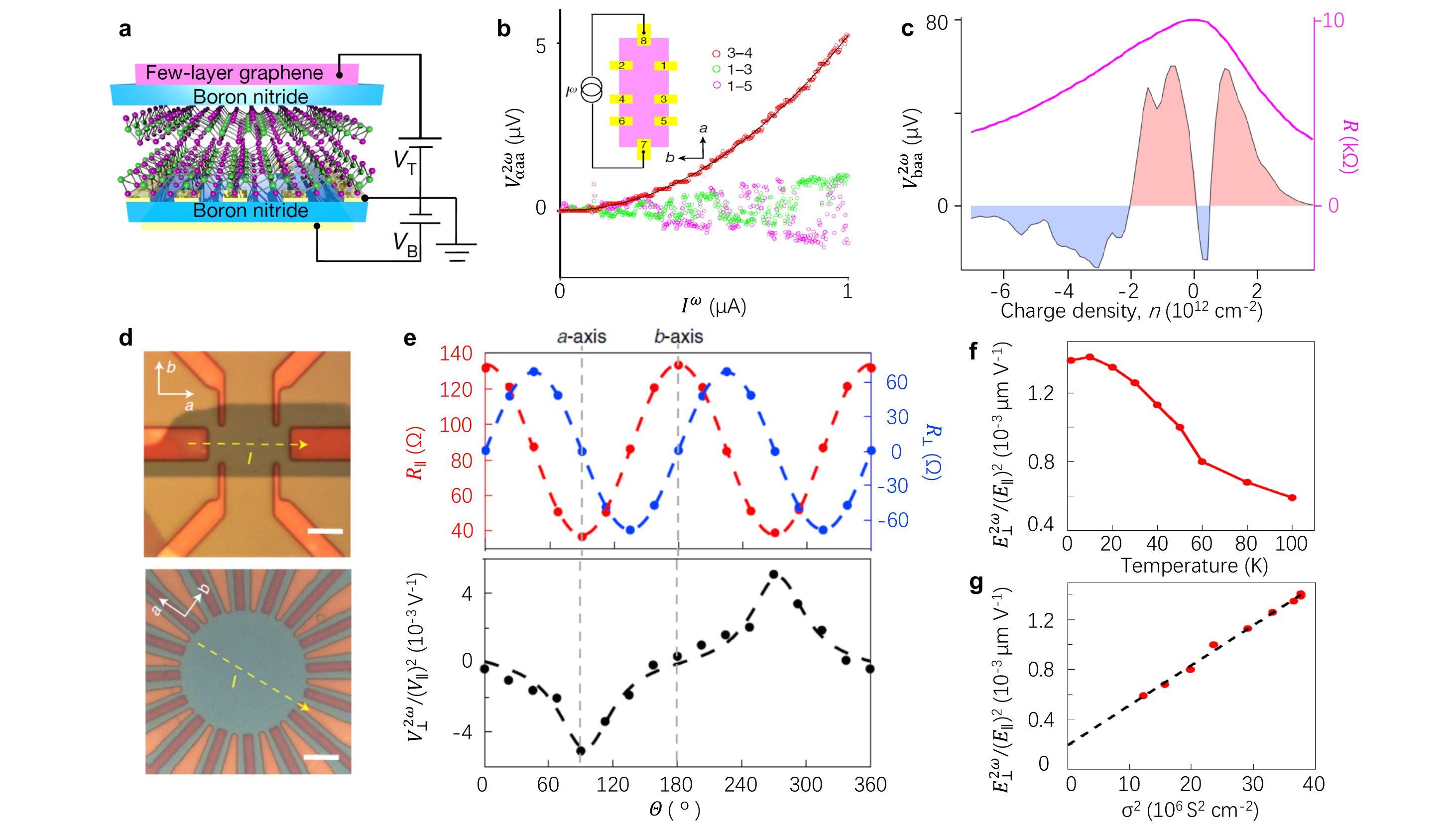}\\
  \caption{\textbf{Experimental studies of the nonlinear Hall effect in WTe$_{2}$}.
  \textbf{a}-\textbf{c} from bilayer samples \cite{Ma19nat}: \textbf{a}, Schematic of an encapsulated, dual-gated, bilayer WTe$_{2}$ device.
  \textbf{b}, The $I$-$V$ relations measured from different electrodes.
  \textbf{c}, $V^{2\omega}_{baa}$ (left, black line) and $R$ (right, magenta line) as a function of charge density. The blue (red) shading highlights negative (positive) $V^{2\omega}_{baa}$.
  \textbf{d}-\textbf{g} from few-layer samples \cite{Kang19nm}:
  \textbf{d}, Optical image of a typical Hall bar (top) and a circular disc device used in this study (bottom). The scale bars are 5 $\mu$m. The crystal orientation was determined by polarised Raman spectroscopy. The WTe$_{2}$ samples are in grey, the Pt electrodes in orange and the Si substrate in dark red-orange for the disc device. Dashed yellow lines indicate the current direction.
  \textbf{e}, $R_{\parallel}$ and $R_{\perp}$ (top) and the nonlinear Hall signal (bottom) of a disc device as a function of $\theta$.
  \textbf{f},\textbf{g}, Nonlinear Hall signal as a function of temperature (\textbf{f}) and the square of the longitudinal conductivity (\textbf{g}).
  Panels \textbf{a}-\textbf{c} adapted from REF.~\cite{Ma19nat}. Panels \textbf{d}-\textbf{g} adapted from REF.~\cite{Kang19nm}.}\label{Fig:Exp}
\end{figure*}

\section{Experimental measurements}

The nonlinear Hall effect can be measured with a Hall bar  (Fig.~\ref{Fig:nonlinear Hall effect} \textbf{a}). To measure a linear Hall effect while using the lock-in technique to suppress noise, an $ac$ current $I_\omega$ oscillating at a low frequency $\omega$ is injected, and a transverse voltage $V_\omega$ of the same frequency is measured. The measurement of a nonlinear Hall effect follows the same geometry, but the transverse voltage is measured at the double-frequency $V_{2\omega}$ (which is more convenient than the zero-frequency). Theoretically, the voltage is formulated as a transverse current density $J$ in response to the electric field $E$ of the injected current and can be formally expressed as a power series
\begin{equation}\label{Eq:Current}
J_{a}=\sigma_{ab}E_{b}+\chi_{abc}E_{b}E_{c}+\cdots,
\end{equation}
where $a,b,c\in(x,y,z)$, $\sigma_{ab}$ ($a\neq b$) is the linear Hall conductivity, and the leading nonlinear term is characterised by the quadratic conductivity tensor $\chi_{abc}$. Because the frequency is extremely low ($\sim$ 10 to 1000 Hz), the measurement is in equilibrium and differs greatly from nonlinear optics  (where the light frequencies are approximately $10^{14}$ Hz).

As a nonlinear nonreciprocal quantum transport phenomenon \cite{Nagaosa2018NCR}, the nonlinear Hall effect has been predicted in two-dimensional (2D) transition metal dichalcogenides \cite{Sodemann15prl,Low15prb,You18prb,ZhangY18-2dm,Du18prl}, on the surface of topological crystalline insulators \cite{Sodemann15prl} and in three-dimensional Weyl semimetals \cite{Sodemann15prl,Zhang18PRB,Facio18prl}.
The effect requires low-symmetry systems and can thus be realised by symmetry-breaking engineering \cite{Sun2021NRP}.

Experimental observations of the nonlinear Hall effect were first reported in 2D layers of WTe$_{2}$ \cite{Ma19nat,Kang19nm}, a material with time-reversal symmetry. Ma \emph{et al}. \cite{Ma19nat} measured the nonlinear Hall effect in encapsulated and dual-gated bilayer WTe$_{2}$ devices (Fig.~\ref{Fig:Exp} \textbf{a}). By applying an $ac$ electric field with frequency $\omega$ (10-1000 Hz), they measured a double-frequency voltage with a lock-in amplifier at $T=10-100$ K. Their measurements demonstrated clear quadratic $I$-$V$ relations (Fig.~\ref{Fig:Exp} \textbf{b}) and were independent of the driving frequency. They also showed that the sign of the nonlinear Hall effect can be tuned by the carrier density and the out-of-plane electrical displacement field via the top and bottom gate voltages (Fig.~\ref{Fig:Exp} \textbf{c}).
The strong gate-tunability shows a major difference between the linear and nonlinear Hall effects. The linear Hall effect is well known to be related to the Berry curvature \cite{Xiao10rmp}. Compared to the linear Hall effect, the nonlinear Hall effect can probe additionally a physical property called the Berry curvature dipole, which can be regarded as the derivative of the Berry curvature. This gate-tunable sign-change behaviour of nonlinear Hall voltage was explained by an effective model \cite{Du18prl} and first-principle calculation \cite{Ma19nat}, which reveal that the gating in the bilayer WTe$_{2}$ can lead to band inversions, which change the sign of the Berry curvature dipole on the Fermi surface, as well as the intrinsic and disorder contributions to the nonlinear Hall effect. Kang \emph{et al}. measured the nonlinear Hall effect in WTe$_{2}$ devices ranging from 4 to
8 layers \cite{Kang19nm}. In addition to the Hall bar, they performed angle-resolved measurements using circular disc devices with electrodes aligned with the crystal axes (Fig.~\ref{Fig:Exp} \textbf{d}).
The second-harmonic transverse voltage is quadratic with longitudinal voltage for all angles. Unlike the two-fold angular dependence of linear voltages, the nonlinear response shows a one-fold angular dependence (Fig.~\ref{Fig:Exp} \textbf{e}), with the maximum (zero point) occurring when the current is along the $a$-axis ($b$-axis).
They also investigated the variation of the nonlinear Hall conductivity as a function of temperature (Fig.~\ref{Fig:Exp} \textbf{f}) and the square of the longitudinal conductivity (Fig.~\ref{Fig:Exp} \textbf{g}). These observations indicate the existence of various types of contributions to the nonlinear Hall effect.

After the first observations, new experimental studies of the nonlinear Hall effect have established a rapidly growing trend.
The nonlinear Hall effect has also been observed in three-dimensional crystals of WTe$_{2}$ (type-II Weyl semimetal), Dirac semimetal Cd$_{3}$As$_{2}$ \cite{Shvetsov19jetp}, nonmagnetic Weyl-Kondo semimetal Ce$_3$Bi$_4$Pd$_3$ \cite{Dzsaber21PNAS}, strain-engineered monolayer WSe$_2$ \cite{Qin2021CPL}, artificially-corrugated bilayer graphene \cite{Ho2021NE}, small-angle-twisted WSe$_2$ \cite{Huang2020arXiv},  T$_\mathrm{d}$-MoTe$_2$ \cite{Tiwari21NC}, multi-layered organic 2D Dirac fermion system $\alpha$-(BEDT-TTF)$_2$I$_3$ \cite{Kiswandhi21arXiv}, the surface of the topological insulator Bi$_2$Se$_3$ at zero magnetic field \cite{He2021NC} (which is argued to be induced purely by skew-scattering mechanism instead of the Berry curvature dipole) and Weyl semimetal TaIrTe$_4$ \cite{Kumar2021NN} at room temperature (Tab.~\ref{Tab:Exp}).

In addition to the observation in different materials, the applications of the nonlinear Hall effect based on their unique features have also been investigated experimentally. This exciting progress are discussed in Sec.~\ref{Sec:Prospects} together with some related theoretical proposals.

\BOX{
\textbf{Box 1: Time-reversal symmetry, Onsager's reciprocal theorem and Hall effects.}

Time-reversal symmetry is one of the most important discrete symmetries in physics.
It represents the symmetry under the transformation that reverse the arrow of time: $t\rightarrow-t$.
For a generic physical operator $\hat{\mathcal{O}}$, its time-reversal transformation is
\begin{equation}
\hat{\mathcal{T}}\hat{\mathcal{O}}\hat{\mathcal{T}}^{-1}=\lambda_{\mathcal{O}}\hat{\mathcal{O}},
\end{equation}
where $\lambda_{\mathcal{O}}=\pm1$ specify the even/odd nature of the quantity $\hat{\mathcal{O}}$.
For example, the time-reversal leaves the position operator $\hat{\bm{r}}$ unchanged ($\lambda_{\bm{r}}=1$) while change the sign of the momentum operator $\hat{\bm{p}}$ ($\lambda_{\bm{p}}=-1$).
If one further considers the time-reversal transformation of their commutator
$\hat{\mathcal{T}}[\hat{\bm{r}},\hat{\bm{p}}]\hat{\mathcal{T}}^{-1}$, one can obtain that
$\hat{\mathcal{T}}i\hat{\mathcal{T}}^{-1}=-i$,
which means that $\hat{\mathcal{T}}$ is an anti-unitary operator \cite{Shen17book}.
This property together with the fact that spin should flips under the time-reversal ($\lambda_{\bm{s}}=-1$) allow one to construct the time-reversal operator for the spin-1/2 particles as $\hat{\mathcal{T}}=-i\sigma_{y}\hat{\mathcal{K}}$, where $\bm{\sigma}=(\sigma_{x}, \sigma_{y}, \sigma_{z})$ are the Pauli matrices and $\hat{\mathcal{K}}$ is the complex-conjugate operator. For spin-$1/2$ particles, one can explicitly check that $\hat{\mathcal{T}}^{2}=-1$, which gives rise to the Kramer's theorem that guarantees the double degeneracy at the time-reversal invariant points in the Brillouin zone.

Besides the Hamiltonian and physical operators, time-reversal symmetry also constrains the correlation functions.
For example, as the charge current $\bm{J}$ is odd under time-reversal while the electric field $\bm{E}$ is time-reversal even, the current response functions in Eq.~(\ref{Eq:Current}) relate quantities with different time-reversal symmetry.
This fact implies that the generic conductivity should be associated with dissipative irreversible processes such as the Joule heating $Q=\bm{J}\cdot\bm{E}$ in the conductor.
The evolution of the heat results in an increase in the entropy.
When an amount of the heat $dQ=\bm{J}\cdot\bm{E}dV$ is evolved for element volume $dV$, the changing rate of the entropy of the system is
$dS/dt=\int(\bm{J}\cdot\bm{E}/T)dV$ at temperature $T$ \cite{Landau-ECM}, which determines the direction of time.
In this sense, time-reversal symmetry is connected to the irreversibility of the transport processes, and this irreversibility is Hamiltonian-independent and appears only when we consider macroscopic systems with continuous energy spectra \cite{Nagaosa10rmp}.

Time-reversal symmetry also has Hamiltonian-dependent constraints on the microscopic dynamics of the system.
The most important one is the Onsager's reciprocal theorem \cite{Onsager1931PR1,Kubo1957}
\begin{equation}\label{Eq:Onsager}
\mathcal{R}_{\mathcal{O}\mathcal{C}}(\bm{q},\omega,\bm{B})=\lambda_{\mathcal{O}}\lambda_{\mathcal{C}}\mathcal{R}_{\mathcal{C}\mathcal{O}}(-\bm{q},\omega,-\bm{B}),
\end{equation}
where $\bm{B}$ represents the magnetic field, $\mathcal{R}_{\mathcal{O}\mathcal{C}}$ is a generic linear response function of a physical quantity $\hat{\mathcal{O}}$ to the perturbation that conjugates to the quantity $\hat{\mathcal{C}}$ with the wavevector $\bm{q}$ and frequency $\omega$.
As an example, we consider the current-current correlation function $\mathcal{R}_{j_{a}j_{b}}$ that is associated with the linear conductivity tensor $\sigma_{ab}$.
Since $\lambda_{\bm{j}}=-1$, the constraint imposed by the Onsager's reciprocal theorem on the linear conductivity tensor is \cite{Nagaosa10rmp}
\begin{equation}\label{Eq:Onsager-Sigma}
\sigma_{ab}(\bm{q},\omega,\bm{B})=\sigma_{ba}(-\bm{q},\omega,-\bm{B}).
\end{equation}
Hence, the linear conductivity tensor $\sigma_{ab}$ should be symmetric with respect to the output and input current directions ($a$ and $b$) in time-reversal symmetric systems,
while the antisymmetric part is finite only if time-reversal symmetry is broken.
In other words, this relation forbids the appearance of the linear charge Hall effects in time-reversal symmetric systems when the electric field is
along the principal axes.
The situation would be different if we consider the spin-current response $\mathcal{R}_{j^{s}_{a}j_{b}}$ that is associated with the spin Hall effect \cite{Sinova15rmp}.
Because $\lambda_{\bm{j}^{s}}=1$, the spin Hall effect is then allowed in time-reversal symmetric systems.

For the nonlinear Hall effects, the related correlation functions are also nonlinear, which describe the correlation between more than two operators.
The constraint Eq.~\ref{Eq:Onsager} is only about the linear response and thus can not forbid the appearance of the Hall effects in the nonlinear-response regime.
Although the theoretical and experimental discoveries of the nonlinear Hall effects have proved this statement, it is still unknown how time-reversal symmetry would constrain the nonlinear response functions.
The nonlinear and nonequilibrium generalizations of the Onsager's reciprocal theorem have been intensively discussed recently in form of the general fluctuation theorem \cite{Evans93PRL,Evans2008,Esposito09RMP,Campisi11RMP,Morimoto18SR}.
Several versions of nonlinear Onsager's reciprocal theorem have been proposed by using different theoretical approaches \cite{Esposito09RMP,Campisi11RMP,Morimoto18SR}, however, a generic yet explicit expression and its implications on the nonlinear transport phenomena have not been well explored \cite{Nagaosa2018NCR}.
}

\begin{table*}[htb]
 \centering
 \caption{\textbf{Experiments on the nonlinear Hall effect}.}\label{Tab:Exp}
 \begin{ruledtabular}
 \begin{tabular}{ccccccc}
     Materials & Dimension & \makecell[c]{Temperature\\(K)} & \makecell[c]{Input current\\frequency\\(Hz)} & \makecell[c]{Input current\\maximum\\($\mu$A)} & \makecell[c]{Output voltage\\maximum\\($\mu$V)} & \makecell[c]{Carrier density\\(cm$^{-2}$) in 2D\\
     (cm$^{-3}$) in 3D} \\
     \hline
     Bilayer WTe$_{2}$ \cite{Ma19nat} & 2 & 10-100  & 10-1000  & 1  & 200 & $\sim10^{12}$ \\
     Few-layer WTe$_{2}$ \cite{Kang19nm} & 2 & 1.8-100  & 17-137  & 600 & 30 & $\sim10^{13}$ \\
     Strained monolayer WSe$_{2}$ \cite{Qin2021CPL} & 2 & 50-140  & 17.777  & 5  & 20  & $\sim10^{13}$ \\
     Twisted bilayer WSe$_{2}$ \cite{Huang2020arXiv} & 2 & 1.5-30  & 4.579  & 0.04 & 20000 & $\sim10^{12}$ \\
     Corrugated bilayer graphene \cite{Ho2021NE} & 2 & 1.5-15  & 77  & 0.1 & 2  & $\sim10^{12}$ \\
     Bi$_{2}$Se$_{3}$ surface \cite{He2021NC} & 2 & 2-200  & 9-263  & 1500 & 20 & $\sim10^{13}$ \\
     Bulk WTe$_{2}$ \cite{Shvetsov19jetp} & 3 & 1.4-4.2  & 110 & 4000 & 2 & - \\
     Cd$_{3}$As$_{2}$ \cite{Shvetsov19jetp} & 3 & 1.4-4.2  & 110  & 4000 & 1 & $\sim10^{18}$ \\
     Ce$_{3}$Bi$_{4}$Pd$_{3}$ \cite{Dzsaber21PNAS} & 3 & 0.4-4  & dc  & 100  & - & - \\
     TaIrTe$_{4}$ \cite{Kumar2021NN} & 3 & 2-300  & 13.7-213.7  & 600 & 120 & $\sim10^{19\text{-}20}$ \\
     T$_\mathrm{d}$-MoTe$_2$ \cite{Tiwari21NC} & 3 & 2-40  & 17-277  & 5000 & 0.4 & $\sim10^{19\text{-}20}$ \\
     $\alpha$-(BEDT-TTF)$_2$I$_3$ \cite{Kiswandhi21arXiv} & 3 & 4.2-40  & dc  & 2000 & 40 & $\sim10^{17}$ \\
   \end{tabular}
   \end{ruledtabular}
\end{table*}

\section{Theoretical understandings}

\subsection{Geometric nature of intrinsic part}
The nonlinear Hall effect can be induced by intrinsic (geometric) and extrinsic (disorder-induced) contributions.
The intrinsic contribution to the nonlinear Hall effect has a geometric nature because of its  connection to the Berry curvature \cite{Sodemann15prl,Low15prb}. The Berry curvature $\bm{\Omega}$ can be regarded as a magnetic field in parameter space (e.g., momentum space) that describes the bending of parameter spaces, which arises from the geometric structure of quantum eigen states \cite{Berry84,Xiao10rmp,Bohm2003Book}.

The Berry curvature is well known for its role in the linear Hall conductivity
\begin{equation}\label{Eq:anomalous Hall effect}
\sigma^{in}_{ab}=-\frac{e^2}{\hbar}\varepsilon_{abc}\int \frac{d^{n}\bm{k}}{(2\pi)^{n}}\Omega_{c}f_{0},
\end{equation}
where
$\varepsilon_{abc}$ is the Levi-Civita symbol, $n$ is the dimension, and $f_{0}$ is the Fermi distribution. The Berry curvature allows one to calculate the intrinsic anomalous Hall effect \cite{Karplus1954PR,Nagaosa10rmp} and can be quantised to the quantum Hall effect in two dimensions \cite{Klitzing80prl,Thouless82prl}.

For the nonlinear Hall effect, the intrinsic part of the nonlinear Hall conductivity tensor
$\chi^{in}_{abc}$ = $(e^3\tau/4\hbar)$ $(\varepsilon_{abd}D_{cd}+b\leftrightarrow c)$ was found to be related to the Berry curvature dipole \cite{Sodemann15prl,Low15prb}
\begin{eqnarray}\label{Eq:BCD}
D_{ab}&=&-\int \frac{d^{n}\bm{k}}{(2\pi)^{n}} \Omega_{b}\partial_{a}f_{0}
=\int \frac{d^{n}\bm{k}}{(2\pi)^{n}} (\partial_{a}\Omega_{b})f_{0},
\end{eqnarray}
where $\tau$ is the relaxation time, $\partial_{a}\equiv\partial/\partial k_{a}$.
In a time-reversal invariant system, the Berry curvature is odd in momentum space \cite{Xiao10rmp}, so a nonzero linear Hall effect requires time-reversal symmetry to be broken, according to Eq.~(\ref{Eq:anomalous Hall effect}). In contrast, the nonlinear Hall effect requires inversion symmetry to be broken, according to Eq.~(\ref{Eq:BCD}).

\begin{figure*}[htpb]
\centering
  \includegraphics[width=0.98\textwidth]{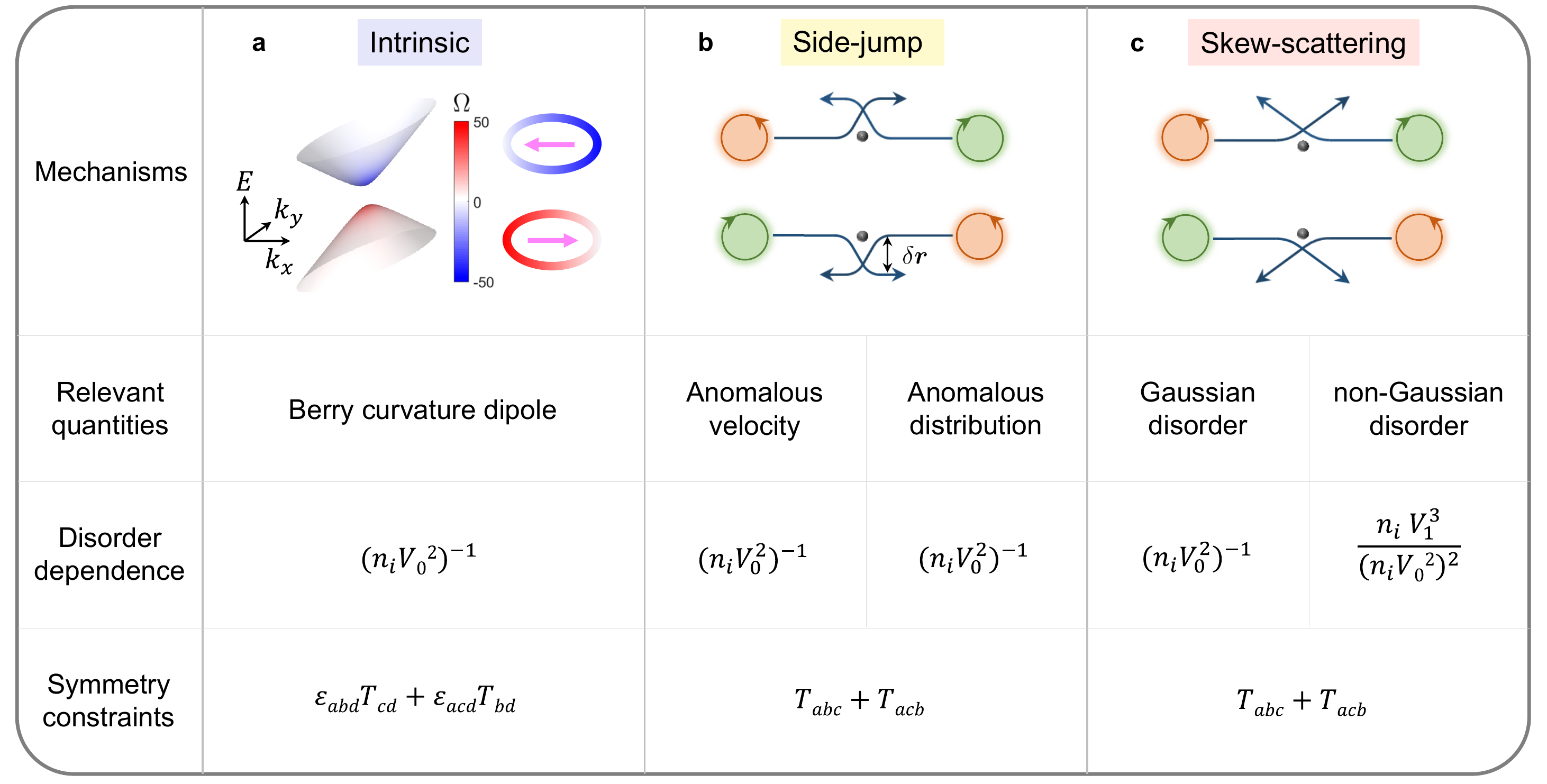}\\
  \caption{\textbf{Known mechanisms of the nonlinear Hall effect }.
  \textbf{a}, Intrinsic Berry curvature dipole mechanism illustrated with the help of the 2D tilted massive Dirac model. Colour bar shows the value of the Berry curvatures $\Omega$. The tilt leads to imbalanced Berry curvature and group velocity, and their product gives a non-zero Berry curvature dipole (purple arrows). The conduction and valence bands have opposite Berry curvature dipoles.
  \textbf{b}, Side-jump mechanism. The motion of self-rotating wave packets (green and orange for clockwise and anticlockwise self-rotating wave packets, respectively) is deflected in opposite directions by the opposite electric fields experienced upon approaching, and leaving an impurity (grey) with a coordinate shift $\delta\bm{r}$.
  \textbf{c}, Skew-scattering mechanism \cite{Isobe20sa}. Asymmetric scattering due to the effective spin-orbit coupling of wave packets or impurity. Two wave packets moving in opposite directions produce zero net current. If the two wave packets self-rotate in opposite directions, both the side-jump and skew-scattering produce net current in the perpendicular direction. $n_i$ is the impurity density, and $V^{2}_{0}$ and $V^{3}_{1}$ are the Gaussian and non-Gaussian disorder correlation strengths, respectively. $\varepsilon_{abc}$ is the Levi-Civita symbol. $T_{ab}$ and $T_{abc}$ stand for rank-two and rank-three tensors, respectively.
  Panel \textbf{c} adapted from REF.~\cite{Isobe20sa}.}\label{Fig:Dirac}
\end{figure*}

To meet the symmetry requirements, the minimal model for the nonlinear Hall effect is a tilted 2D massive Dirac model
\begin{eqnarray}\label{Eq:TiltedDirac}
\hat{\mathcal{H}}=tk_{x}+v(k_{y}\sigma_{x}+k_{x}\sigma_{y})+(m/2-\alpha k^{2})\sigma_{z},
\end{eqnarray}
where $t$ breaks the inversion symmetry along the $x$ direction by tilting the Dirac cone, ($\sigma_{x}$, $\sigma_{y}$ and $\sigma_{z}$) are the Pauli matrices, $m$ is the energy gap, and $\alpha$ regulates topological properties as $\bm{k}\rightarrow\infty$ \cite{Shen17book}. The model hosts a strong distribution of the Berry curvature near the band edge (Fig.~\ref{Fig:Dirac} \textbf{a}). The time-reversal counterpart of the model makes an equal contribution to the nonlinear Hall effect.  With the model, significant enhancement of the Berry curvature dipole is found near the band anti-crossings, and the dipole changes sign when crossing the gap \cite{Du18prl,Facio18prl} and thus is highly tunable via strain engineering \cite{Xu2018NP,Son2019PRL,Battilomo2019PRL,Chen2019spin,ZhouT20prap,Singh20PRL,Xiao20PRApp} and the twisted bilayer design \cite{Law2020arXivWSe,Law2020arXivTBG,Pantaleon2020PRB,Weng21arXiv}.

\subsection{Disorder-induced extrinsic part}
Disorder is a major contribution to the linear Hall effects, such as the localization in the quantum Hall effects \cite{Prange12book,Chang13sci}. The long debate on the intrinsic (Berry curvature) and extrinsic (disorder) origins of the anomalous Hall effect \cite{Karplus1954PR,Smit1955,Smit1958,Berger1970PRB,Nagaosa10rmp,Bruno2001PRB,Sinitsyn07jpcm,Sinitsyn07prb} reveals that both parts are equally important \cite{Tian09PRL,Hou15PRL,Yue16JPSJ}. Disorder has greater importance in the nonlinear Hall effect, in which the Fermi energy crosses the energy bands so that disorder scattering is inevitable and occurs even in the leading order.

One approach to address the disorder contributions is the semiclassical Boltzmann equation
\cite{Ashcroft1976}
\begin{equation}\label{Eq:Boltzmann}
\frac{\partial f_{l}}{\partial t}+\dot{\bm{k}}\cdot\frac{\partial f_{l}}{\partial\bm{k}}
=-\sum_{l'}(W_{l'l}f_{l}-W_{ll'}f_{l'}),
\end{equation}
where $W_{ll'}$ is the disorder-averaged scattering rate from state $l'$ to $l$. In the Boltzmann formalism, the disorder-induced extrinsic part can be classified into side-jump and skew-scattering mechanisms.

The side-jump mechanism results from the transverse displacement $\delta\bm{r}_{ll'}$ of quasiparticles \cite{Berger1970PRB,Smit1955,Smit1958} (Fig.~\ref{Fig:Dirac} \textbf{b}). In an electric field, $\delta\bm{r}_{ll'}$ leads to an energy shift that modifies the scattering rate \cite{Sinitsyn07jpcm,Du19nc}. This modification gives an anomalous distribution that contributes to the Hall conductivity. In addition, the accumulation of $\delta\bm{r}_{ll'}$ after many scattering events can produce an effective transverse velocity (the side-jump velocity) that also contributes to the Hall conductivity.

The skew-scattering mechanism \cite{Smit1955,Smit1958} originates from transverse asymmetric scattering (Fig.~\ref{Fig:Dirac} \textbf{c}). The scattering rate can be written as
$W_{ll'}=W^{(2)}_{ll'}+W^{(3)}_{ll'}+W^{(4)}_{ll'}+\cdots$ \cite{Sinitsyn07jpcm, Du19nc}, with $W^{(2)}_{ll'}\sim\langle V_{ll'}V_{l'l}\rangle$, $W^{(3)}_{ll'}\sim\langle V_{ll'}V_{l'l''}V_{l''l}\rangle$,
$W^{(4)}_{ll'}\sim\langle V_{ll'}V_{l'l''}V_{l''l'''}V_{l'''l}\rangle$, where $\langle ... \rangle$ represents the disorder average of the scattering matrix elements $V_{ll'}$. $W^{(2)}_{ll'}$ is purely symmetric, so the skew-scattering contribution comes from the antisymmetric parts of $W^{(3)}_{ll'}$ (non-Gaussian disorder distribution) and $W^{(4)}_{ll'}$ (Gaussian disorder distribution) \cite{Sinitsyn07jpcm}.

The disorder part has connections to geometry. In a limit, both disorder mechanisms can be related to the Pancharatnam phase $\Phi_{ll'l''}$=$\arg(\langle u_{l}|u_{l'}\rangle$$\langle u_{l'}|u_{l''}\rangle$$\langle u_{l''}|u_{l}\rangle)$ \cite{Pancharatnam1956,Sinitsyn2006PRB}, a special Berry phase where $\arg(...)$ is the complex angle. In addition, the transverse displacement $\delta\bm{r}_{ll'}=\bm{\mathcal{A}}_{l}-\bm{\mathcal{A}}_{l'} -(\partial\bm{k}+\partial/\partial\bm{k}')\arg(V_{ll'})$ \cite{Sinitsyn2006PRB}, where
$\bm{\mathcal{A}}_{l}=\langle u_{l}|i\partial u_{l}/\partial\bm{k}\rangle$ is the Berry connection. These imply a general form of the Berry curvature with disorder and interactions \cite{Resta21arXiv}.

Attempts have been made to go beyond the semiclassical Boltzmann equation \cite{Nandy19prb,Xiao19prb,Du2020arXiv,Gao2020PRB,Levchenko2021arXiv}. The quantum kinetic equation theory has shown side-jump contributions that have no semiclassical counterpart \cite{Xiao19prb}. A quantum theory based on the Feynman diagram technique has also been constructed, in which multiple new terms have been found \cite{Du2020arXiv}.

The coexistence of the intrinsic and extrinsic mechanisms makes the research of nonlinear Hall effect more interesting yet more complex.
A series of follow-up questions then become significant, such as how to distinguish between the intrinsic and extrinsic contributions, and can the disorder effects even be dominant.
In the linear response it is known that scalar disorder produces a contribution to the anomalous Hall effect that reduces and sometimes exactly cancels the Berry curvature contribution \cite{Culcer17PRB}.
Thus, the competition between the intrinsic and extrinsic contributions can be expected to be important as well.
In the following two subsections we will discuss two solutions to these questions, the scaling law and symmetry analysis of the nonlinear Hall effect.

\subsection{Scaling law}

The scaling law of the Hall effects refers to a relation
between the transverse Hall conductivity $\sigma_H$ and the longitudinal conductivity $\sigma_L$, such as $\sigma_H \sim (\sigma_L)^\alpha$, with $\alpha$ as the scaling factor. Different scaling behaviours allow one to distinguish various transport regimes (high-conductivity, good-metal, and bad-metal-hopping regimes) \cite{Nagaosa10rmp} and different contributions to the anomalous \cite{Tian09PRL,Hou15PRL} and nonlinear Hall effects \cite{Kang19nm,Du19nc}.

The scaling behaviours of the nonlinear Hall effect have undergone preliminary investigation in few-layer WTe$_{2}$ samples (Fig.~\ref{Fig:Exp} \textbf{f} and \textbf{g}), and the results imply that another contribution exists in addition to the Berry curvature dipole \cite{Kang19nm}. A scaling law has been theoretically proposed \cite{Du19nc}, in which a detailed experimental proposal is also described.
Later, an experimental verification of the scaling behaviour was performed in T$_\mathrm{d}$-MoTe$_2$ \cite{Tiwari21NC}, where it was found that the skew-scattering mechanism dominates at higher temperatures and another scattering process with temperature dependent scaling parameters at low temperatures.
The establishment of scaling laws would lead to a more comprehensive understanding of the nonlinear Hall effect.

\subsection{Symmetry analysis}
Symmetry analysis helps to determine nonzero elements in the nonlinear conductivity tensor. $\bm{J}$ and $\bm{E}$ are odd under inversion, so $\chi$ also changes sign under inversion, and the quadratic nonlinear response requires breaking inversion symmetry, regardless of the intrinsic or extrinsic mechanisms. Additionally, a mirror reflection $\mathcal{M}_{a}$ with respect to the $b$-$c$ plane demands all the $\chi$ with odd numbers of $a$ in subscript to vanish. Similar analyses apply to all other symmetry operations.

Symmetry analysis may also help to distinguish the intrinsic and extrinsic contributions. The nonlinear Hall conductivity tensor can be decomposed as $\chi=\chi^{in}+\chi^{ex}$, where the intrinsic and extrinsic part are
\begin{eqnarray}\label{Eq:From}
\chi^{ex}_{abc}\sim&T_{abc}+T_{acb}, \ \
\chi^{in}_{abc}\sim&\varepsilon_{abd}T_{cd}+\varepsilon_{acd}T_{bd},
\end{eqnarray}
where $T_{ab}$ and $T_{abc}$ stand for rank-two and rank-three tensors, respectively. The symmetry properties of the intrinsic contribution originate from the Berry curvature dipole \cite{Sodemann15prl}, which has been explored for all 32 point groups. A further investigation showed that $\chi^{ex}$ has more nonzero elements than $\chi^{in}$ due to disorder, such as $C_{3}$, $C_{3h}$, $C_{3v}$, $D_{3h}$ and $D_{3}$ in two dimensions, and even pure disorder-induced nonlinear Hall effects in $T$, $T_{d}$, $C_{3h}$ and $D_{3h}$ in three dimensions \cite{Du2020arXiv}.

\section{Prospects}\label{Sec:Prospects}

\subsection{Spectroscopic applications in quantum matter}
The nonlinear Hall effect is an unconventional response that is sensitive to the breaking of discrete and crystal symmetries and thus can be used to probe phases or phase transitions induced by the spontaneous breaking of spatial symmetry. The nonlinear Hall effect is more sensitive to the low-energy modes in metallic systems than are optical approaches, such as second-harmonic generation \cite{Zhao16np,Zhao17np}, so it can be used to detect ferroelectric metals \cite{Xiao20PRB} and interaction-driven symmetry breaking in a metallic Dirac system \cite{Habib20prr}.
Because the Berry curvature dipole diverges near a topological transition, it can be used to probe the topological quantum critical points \cite{Facio18prl,Du18prl}.
The strong dependence of the Berry curvature dipole on the N\'{e}el vector orientation provides a
new detection scheme for the N\'{e}el vector in antiferromagnets based on the nonlinear Hall effect \cite{Shao20prl}.
Moreover, because measurement of the nonlinear Hall effect does not depend on light sources and can be performed under some extreme conditions (such as under strong magnetic fields and ultrahigh pressures), it can be used to detect the spatial symmetry-related physics in systems that are inaccessible to optical methods.

\begin{widetext}

\begin{figure}[htpb]
\centering
\includegraphics[width=0.98\textwidth]{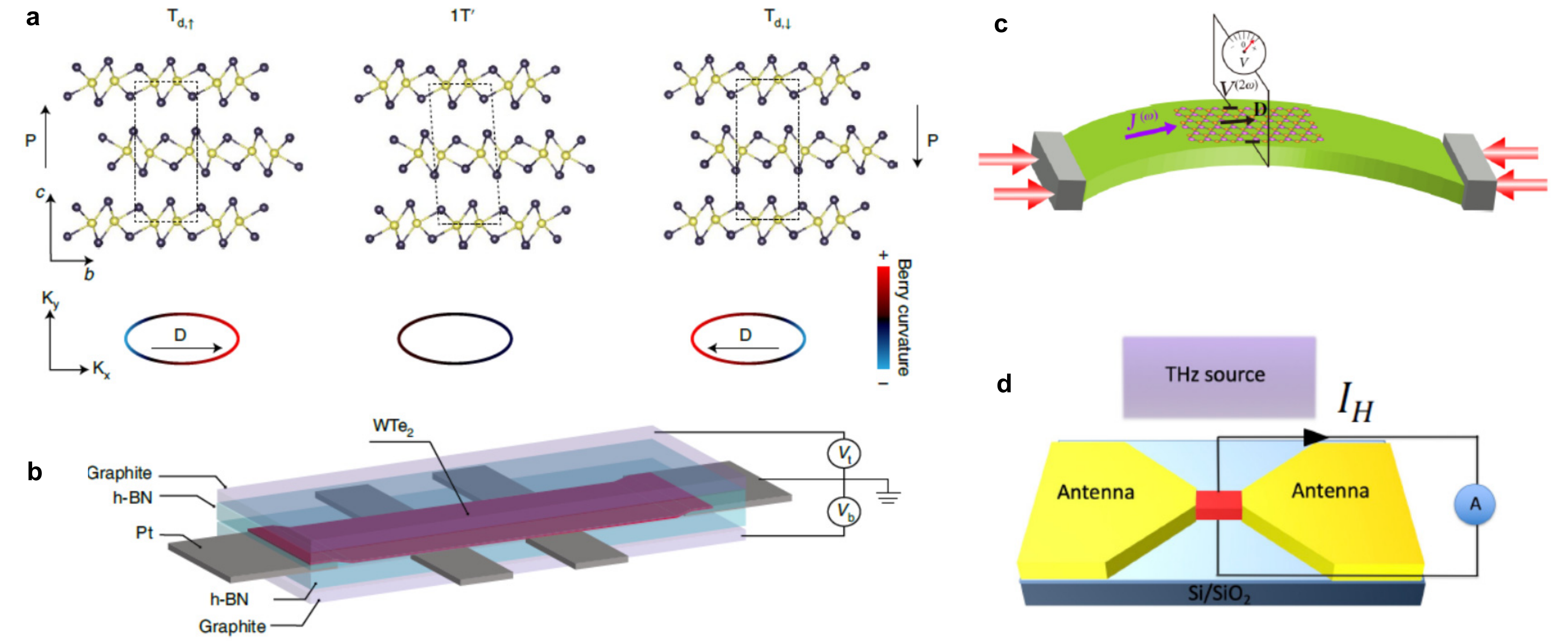}
\caption{\textbf{Device applications of the nonlinear Hall effect}.
\textbf{a}-\textbf{b}, Berry curvature memory device \cite{Xiao2020NP}: \textbf{a}, Side view (the b-c plane) of unit cell shows possible stacking orders in WTe$_2$ (monoclinic 1T$'$, polar orthorhombic T$_{d,\uparrow}$ or T$_{d,\downarrow}$) and schematics of their Berry curvature distributions in momentum space.
P and D represent the spontaneous polarisation and the Berry curvature dipole, respectively. Yellow and black spheres represent W and Te atoms, respectively. \textbf{b}, Schematic of a dual-gate hexagonal boron nitride (h-BN)-capped WTe$_2$ device, in which electrically driven stacking transitions can be realised to enable layer-parity selective memory behaviours of the Berry curvature and Berry curvature dipole with the nonlinear Hall effect.\textbf{c}, Metallic piezoelectric-like device based on the nonlinear Hall effect \cite{Xiao20PRApp}. In the absence of strain, no nonlinear Hall effect exists. In the presence of a uniaxial strain, a charge current $J(\omega)$ leads to a finite nonlinear Hall voltage $V(2\omega)$ as a result of emerging D. \textbf{d}, Photodetection device based on the nonlinear Hall effect \cite{Isobe20sa,Fu2021arXiv}. The oscillating terahertz (THz) field collected by the bowtie antenna induces a direct current in the transverse direction via an intrinsic or extrinsic second-order response of the device. Panels \textbf{a}-\textbf{b} adapted from REF.~\cite{Xiao2020NP}. Panel \textbf{c} adapted from REF.~\cite{Xiao20PRApp}. Panel \textbf{d} adapted from REF.~\cite{Fu2021arXiv}.
}
\label{Fig:Dev}
\end{figure}

\end{widetext}

\subsection{Device applications}
In few-layer WTe$_2$, the nonlinear Hall effect can be used as a reading mechanism of the Berry curvature memory determined by layer number parity (Fig.~\ref{Fig:Dev} \textbf{a} and \textbf{b}) \cite{Xiao2020NP}.
A strain-tunable Berry curvature dipole allows the design of piezoelectric-like devices (Fig.~\ref{Fig:Dev} \textbf{c}) such as strain sensors \cite{Xiao20PRApp}.
The nonlinear Hall effect may provide an alternative approach to current rectification without the use of semiconductor junctions \cite{Isobe20sa}.
According to theoretical calculations, large and tunable rectification can be achieved in graphene
multilayers or transition metal dichalcogenides via the disorder-induced skew-scattering mechanism \cite{Du19nc}, which can be applied to low-power energy harvesters
and terahertz detection \cite{Isobe20sa,Kim17PRB,Bhalla20PRL}
A method for broadband long-wavelength photodetection (Fig.~\ref{Fig:Dev} \textbf{d}) based on the Berry curvature dipole induced nonlinear Hall effect has also been proposed \cite{Fu2021arXiv}.
The Weyl semimetal NbP is identified as a good candidate for terahertz photodetection, with great
current responsivity reaching $\sim$1A/W without external bias \cite{Fu2021arXiv}.
These exciting results on many fronts encourage the future development of terahertz/infrared technologies based on the nonlinear Hall effect.

\subsection{Generalisations to other nonlinear transport effects}
Inspired by the concept of the nonlinear Hall effect, several related nonlinear transport effects have also been proposed.
In the gyrotropic Hall effect, an $ac$ Hall response is induced by injecting a $dc$ current into metals with gyrotropic point groups \cite{Konig19prb}.
A nonlinear spin Hall effect \cite{Hamamoto2017PRB,Araki2018SR} was also proposed as a spin current version of the nonlinear Hall effect.
By replacing one or both driving fields with a temperature gradient $\nabla T$, nonlinear electric transport can be generated to the nonlinear thermoelectric and thermal effects \cite{Nakai2019PRB,Su19prb,Zeng19prb,Zeng2020PRR}, including the nonlinear Nernst effect \cite{Su19prb,Zeng19prb} and the nonlinear anomalous thermal Hall effect \cite{Zeng2020PRR}.
By replacing one of the driving electric fields with the gradient $\bm{E}=(1/e)\partial U/\partial\bm{r}$ of an electrostatic potential $U(\bm{r})$, one can generate the Magnus Hall effect, a self-rotating Bloch electron wave packet that moves under an electric field and develops an anomalous velocity in the transverse direction \cite{Papaj19prl}.
Its thermoelectric and thermal generalisations are the Magnus Nernst effect and Magnus thermal Hall effect \cite{Mandal2020PRB}.
The nonlinear Hall effect can also be generalised to the hydrodynamic regime \cite{Toshio2020PRR}, which describes the anomalous nonlinear/nonlocal responses under nonuniform hydrodynamic variables.
A generalization to higher-order nonlinear Hall response has been proposed, which highlights the importance of the Berry curvature multipoles (the higher-order moments of the Berry curvature near the Fermi surface) \cite{Law2020arXivHigher}.

Although the nonlinear Hall effect can occur in time-reversal symmetric systems, its generalisations to cases in which time-reversal symmetry is broken have also attracted great attention.
In the presence of a magnetic field, the experimental setup of the nonlinear Hall effect would become identical to nonlinear magnetoelectric resistance \cite{He2018NP,He2018PRL,He2019NC} and the nonlinear planar Hall effect \cite{He2019PRL}, whose dominant contribution may not come from the Berry curvature dipole, side-jump or skew-scattering mechanism, but rather from a trivial mechanism (i.e., that which dominates the nonlinear longitudinal conductivity) \cite{Zhang2018Spin,Fert2020PRL}.
For topological semimetals in magnetic fields, nonlinear Hall effects can be further incorporated with the anomaly-induced transport phenomena \cite{Zyuzin2018PRB,zeng2020arXivChiral,Zhang2021PRB}.
Another important generalisation would be to magnetic materials. Various nonlinear transport phenomena have been observed experimentally in the magnetic Weyl semimetal Co$_3$Sn$_2$S$_2$ \cite{Deviatov2020JETP}. The nonlinear Hall effect seen in odd-parity magnetic multipole systems has also been addressed theoretically \cite{Watanabe2021PRR}.
More importantly, nearly all of these generalisations can be implemented in an experimental manner by making simple adjustments to the experimental setups. Therefore, the nonlinear Hall effect has generated a whole new branch of the Hall family, with many outstanding directions for future exploration.


\section{Conclusions}
The nonlinear Hall effect differs from nonlinear optics in many aspects. In nonlinear optical effects \cite{Boyd1992Book}, such as the photovoltaic effect and second-harmonic generation, the frequency of light is approximately $10^{14}$ Hz \cite{Xu2018NP}, which is sufficient to trigger interband transitions in semiconductors.
In contrast, the frequency in nonlinear Hall measurements is too low to trigger interband transitions, which confines the physics near the Fermi surface.
Another important energy scale is the relaxation time $\tau$, which can be used to define the Boltzmann limit ($\omega\tau\gg1$) and the diffusive limit ($\omega\tau\ll1$).
The low frequency (approximately $10$-$1000$ Hz) in the experiments \cite{Ma19nat,Kang19nm} confines the system within the diffusive limit, where the importance of interband coherence effects and disorder effects is self-evident and the conductivity of the system rarely shows much frequency dependence \cite{Flensberg95PRB,Kamenev95PRB}.

The nonlinear generalisation of the Hall effects drives the study of quantum transport and topological physics to the nonlinear-response regime, thus providing many interesting prospects for future research.
The concept of the nonlinear Hall effect has already led to a wealth of proposed and experimentally observed phenomena, each with distinct characteristics.
One important research direction on the theoretical front will be to acquire a quantitative understanding of the underlying mechanisms and to further explore the potential applications of each of the nonlinear Hall effects.
For example, most of the known experiments include analysis of the Berry curvature dipole of the electronic band structure, whilst a quantitative estimation of the disorder/interaction effects is usually absent.
The main problem is that no consistent theory of the nonlinear Hall effect exists beyond the Berry curvature dipole and semiclassical description of the disorder effects. When magnetism and heat transport are further considered, this problem becomes even more challenging. On the experimental front, it would be desirable to identify more new nonlinear Hall effects and more materials. This would allow researchers to explore applications of the newly discovered nonlinear Hall effects and to invent new technologies based on these.
Because the geometry-related higher-order response is generic, the extension of the topological nonlinear response to other areas of physics is eagerly awaited.

\begin{acknowledgments}
We thank helpful discussions with Huimei Liu, Suyang Xu, Tse-Ming Chen, Andhika Kiswandhi and Silke Buehler-Paschen.
This work was supported by the National Natural Science Foundation of China (11534001, 12004157, and 11925402), the Strategic Priority Research Program of Chinese Academy of Sciences (Grant No. XDB28000000), Guangdong (2016ZT06D348), the National Key R \& D Program (2016YFA0301700), Shenzhen High-level Special Fund (No. G02206304, G02206404), and the Science, Technology and Innovation Commission of Shenzhen Municipality (ZDSYS20170303165926217, JCYJ20170412152620376).
\end{acknowledgments}


\begin{thebibliography}{100}
\expandafter\ifx\csname url\endcsname\relax
  \def\url#1{\texttt{#1}}\fi
\expandafter\ifx\csname urlprefix\endcsname\relax\def\urlprefix{URL }\fi
\providecommand{\bibinfo}[2]{#2}
\providecommand{\eprint}[2][]{\url{#2}}

\bibitem{Hall1879}
\bibinfo{author}{Hall, E.~H.} \emph{et~al.}
\newblock \bibinfo{title}{On a new action of the magnet on electric currents}.
\newblock \emph{\bibinfo{journal}{American Journal of Mathematics}}
  \textbf{\bibinfo{volume}{2}}, \bibinfo{pages}{287--292}
  (\bibinfo{year}{1879}).
\newblock \urlprefix\url{https://www.nature.com/articles/021361a0}.

\bibitem{Hall1881}
\bibinfo{author}{Hall, E.~H.}
\newblock \bibinfo{title}{{XVIII.} {On} the rotational coefficient in nickel
  and cobalt}.
\newblock \emph{\bibinfo{journal}{The London, Edinburgh, and Dublin
  Philosophical Magazine and Journal of Science}}
  \textbf{\bibinfo{volume}{12}}, \bibinfo{pages}{157--172}
  (\bibinfo{year}{1881}).
\newblock
  \urlprefix\url{https://www.tandfonline.com/doi/abs/10.1080/14786448108627086?journalCode=tphm16}.

\bibitem{Klitzing80prl}
\bibinfo{author}{Klitzing, K.~v.}, \bibinfo{author}{Dorda, G.} \&
  \bibinfo{author}{Pepper, M.}
\newblock \bibinfo{title}{New method for high-accuracy determination of the
  fine-structure constant based on quantized {Hall} resistance}.
\newblock \emph{\bibinfo{journal}{Phys. Rev. Lett.}}
  \textbf{\bibinfo{volume}{45}}, \bibinfo{pages}{494--497}
  (\bibinfo{year}{1980}).
\newblock \urlprefix\url{http://link.aps.org/doi/10.1103/PhysRevLett.45.494}.

\bibitem{Tsui82prl}
\bibinfo{author}{Tsui, D.~C.}, \bibinfo{author}{Stormer, H.~L.} \&
  \bibinfo{author}{Gossard, A.~C.}
\newblock \bibinfo{title}{Two-dimensional magnetotransport in the extreme
  quantum limit}.
\newblock \emph{\bibinfo{journal}{Phys. Rev. Lett.}}
  \textbf{\bibinfo{volume}{48}}, \bibinfo{pages}{1559--1562}
  (\bibinfo{year}{1982}).
\newblock \urlprefix\url{https://link.aps.org/doi/10.1103/PhysRevLett.48.1559}.

\bibitem{Chang13sci}
\bibinfo{author}{Chang, C.-Z.} \emph{et~al.}
\newblock \bibinfo{title}{Experimental observation of the quantum anomalous
  {Hall} effect in a magnetic topological insulator}.
\newblock \emph{\bibinfo{journal}{Science}} \textbf{\bibinfo{volume}{340}},
  \bibinfo{pages}{167--170} (\bibinfo{year}{2013}).
\newblock
  \urlprefix\url{http://www.sciencemag.org/content/340/6129/167.abstract}.

\bibitem{Prange12book}
\bibinfo{author}{Cage, M.~E.} \emph{et~al.}
\newblock \emph{\bibinfo{title}{The quantum {Hall} effect}}
  (\bibinfo{publisher}{Springer Science \& Business Media},
  \bibinfo{year}{2012}).

\bibitem{Nagaosa10rmp}
\bibinfo{author}{Nagaosa, N.}, \bibinfo{author}{Sinova, J.},
  \bibinfo{author}{Onoda, S.}, \bibinfo{author}{MacDonald, A.~H.} \&
  \bibinfo{author}{Ong, N.~P.}
\newblock \bibinfo{title}{Anomalous {Hall} effect}.
\newblock \emph{\bibinfo{journal}{Rev. Mod. Phys.}}
  \textbf{\bibinfo{volume}{82}}, \bibinfo{pages}{1539--1592}
  (\bibinfo{year}{2010}).
\newblock \urlprefix\url{https://link.aps.org/doi/10.1103/RevModPhys.82.1539}.

\bibitem{Nakatsuji2015Nature}
\bibinfo{author}{Nakatsuji, S.}, \bibinfo{author}{Kiyohara, N.} \&
  \bibinfo{author}{Higo, T.}
\newblock \bibinfo{title}{Large anomalous {Hall} effect in a non-collinear
  antiferromagnet at room temperature}.
\newblock \emph{\bibinfo{journal}{Nature}} \textbf{\bibinfo{volume}{527}},
  \bibinfo{pages}{212--215} (\bibinfo{year}{2015}).
\newblock \urlprefix\url{https://www.nature.com/articles/nature15723}.

\bibitem{Machida2010Nature}
\bibinfo{author}{Machida, Y.}, \bibinfo{author}{Nakatsuji, S.},
  \bibinfo{author}{Onoda, S.}, \bibinfo{author}{Tayama, T.} \&
  \bibinfo{author}{Sakakibara, T.}
\newblock \bibinfo{title}{Time-reversal symmetry breaking and spontaneous hall
  effect without magnetic dipole order}.
\newblock \emph{\bibinfo{journal}{Nature}} \textbf{\bibinfo{volume}{463}},
  \bibinfo{pages}{210--213} (\bibinfo{year}{2010}).
\newblock \urlprefix\url{https://www.nature.com/articles/nature08680}.

\bibitem{Yasuda16np}
\bibinfo{author}{Yasuda, K.} \emph{et~al.}
\newblock \bibinfo{title}{Geometric {Hall} effects in topological insulator
  heterostructures}.
\newblock \emph{\bibinfo{journal}{Nat. Phys.}} \textbf{\bibinfo{volume}{12}},
  \bibinfo{pages}{555--559} (\bibinfo{year}{2016}).
\newblock
  \urlprefix\url{http://www.nature.com/nphys/journal/v12/n6/full/nphys3671.html}.

\bibitem{Onsager1931PR1}
\bibinfo{author}{Onsager, L.}
\newblock \bibinfo{title}{Reciprocal relations in irreversible processes. {I}.}
\newblock \emph{\bibinfo{journal}{Phys. Rev.}} \textbf{\bibinfo{volume}{37}},
  \bibinfo{pages}{405--426} (\bibinfo{year}{1931}).
\newblock \urlprefix\url{https://link.aps.org/doi/10.1103/PhysRev.37.405}.

\bibitem{Sinova15rmp}
\bibinfo{author}{Sinova, J.}, \bibinfo{author}{Valenzuela, S.~O.},
  \bibinfo{author}{Wunderlich, J.}, \bibinfo{author}{Back, C.~H.} \&
  \bibinfo{author}{Jungwirth, T.}
\newblock \bibinfo{title}{Spin {Hall} effects}.
\newblock \emph{\bibinfo{journal}{Rev. Mod. Phys.}}
  \textbf{\bibinfo{volume}{87}}, \bibinfo{pages}{1213--1260}
  (\bibinfo{year}{2015}).
\newblock \urlprefix\url{https://link.aps.org/doi/10.1103/RevModPhys.87.1213}.

\bibitem{Xiao2007PRL}
\bibinfo{author}{Xiao, D.}, \bibinfo{author}{Yao, W.} \& \bibinfo{author}{Niu,
  Q.}
\newblock \bibinfo{title}{Valley-contrasting physics in graphene: Magnetic
  moment and topological transport}.
\newblock \emph{\bibinfo{journal}{Phys. Rev. Lett.}}
  \textbf{\bibinfo{volume}{99}}, \bibinfo{pages}{236809}
  (\bibinfo{year}{2007}).
\newblock
  \urlprefix\url{https://link.aps.org/doi/10.1103/PhysRevLett.99.236809}.

\bibitem{Yao2008PRB}
\bibinfo{author}{Yao, W.}, \bibinfo{author}{Xiao, D.} \& \bibinfo{author}{Niu,
  Q.}
\newblock \bibinfo{title}{Valley-dependent optoelectronics from inversion
  symmetry breaking}.
\newblock \emph{\bibinfo{journal}{Phys. Rev. B}} \textbf{\bibinfo{volume}{77}},
  \bibinfo{pages}{235406} (\bibinfo{year}{2008}).
\newblock \urlprefix\url{https://link.aps.org/doi/10.1103/PhysRevB.77.235406}.

\bibitem{Xiao2012PRL}
\bibinfo{author}{Xiao, D.}, \bibinfo{author}{Liu, G.-B.},
  \bibinfo{author}{Feng, W.}, \bibinfo{author}{Xu, X.} \& \bibinfo{author}{Yao,
  W.}
\newblock \bibinfo{title}{Coupled spin and valley physics in monolayers of
  ${\mathrm{mos}}_{2}$ and other group-vi dichalcogenides}.
\newblock \emph{\bibinfo{journal}{Phys. Rev. Lett.}}
  \textbf{\bibinfo{volume}{108}}, \bibinfo{pages}{196802}
  (\bibinfo{year}{2012}).
\newblock
  \urlprefix\url{https://link.aps.org/doi/10.1103/PhysRevLett.108.196802}.

\bibitem{Mak2014Sci}
\bibinfo{author}{Mak, K.~F.}, \bibinfo{author}{McGill, K.~L.},
  \bibinfo{author}{Park, J.} \& \bibinfo{author}{McEuen, P.~L.}
\newblock \bibinfo{title}{The valley {Hall} effect in {MoS$_{2}$} transistors}.
\newblock \emph{\bibinfo{journal}{Science}} \textbf{\bibinfo{volume}{344}},
  \bibinfo{pages}{1489--1492} (\bibinfo{year}{2014}).
\newblock
  \urlprefix\url{https://science.sciencemag.org/content/344/6191/1489.abstract}.

\bibitem{Sodemann15prl}
\bibinfo{author}{Sodemann, I.} \& \bibinfo{author}{Fu, L.}
\newblock \bibinfo{title}{Quantum nonlinear {Hall} effect induced by {Berry}
  curvature dipole in time-reversal invariant materials}.
\newblock \emph{\bibinfo{journal}{Phys. Rev. Lett.}}
  \textbf{\bibinfo{volume}{115}}, \bibinfo{pages}{216806}
  (\bibinfo{year}{2015}).
\newblock
  \urlprefix\url{https://link.aps.org/doi/10.1103/PhysRevLett.115.216806}.

\bibitem{Low15prb}
\bibinfo{author}{Low, T.}, \bibinfo{author}{Jiang, Y.} \&
  \bibinfo{author}{Guinea, F.}
\newblock \bibinfo{title}{Topological currents in black phosphorus with broken
  inversion symmetry}.
\newblock \emph{\bibinfo{journal}{Phys. Rev. B}} \textbf{\bibinfo{volume}{92}},
  \bibinfo{pages}{235447} (\bibinfo{year}{2015}).
\newblock \urlprefix\url{https://link.aps.org/doi/10.1103/PhysRevB.92.235447}.

\bibitem{Facio18prl}
\bibinfo{author}{Facio, J.~I.} \emph{et~al.}
\newblock \bibinfo{title}{Strongly enhanced {B}erry dipole at topological phase
  transitions in {BiTeI}}.
\newblock \emph{\bibinfo{journal}{Phys. Rev. Lett.}}
  \textbf{\bibinfo{volume}{121}}, \bibinfo{pages}{246403}
  (\bibinfo{year}{2018}).
\newblock
  \urlprefix\url{https://link.aps.org/doi/10.1103/PhysRevLett.121.246403}.

\bibitem{You18prb}
\bibinfo{author}{You, J.-S.}, \bibinfo{author}{Fang, S.}, \bibinfo{author}{Xu,
  S.-Y.}, \bibinfo{author}{Kaxiras, E.} \& \bibinfo{author}{Low, T.}
\newblock \bibinfo{title}{Berry curvature dipole current in the transition
  metal dichalcogenides family}.
\newblock \emph{\bibinfo{journal}{Phys. Rev. B}} \textbf{\bibinfo{volume}{98}},
  \bibinfo{pages}{121109} (\bibinfo{year}{2018}).
\newblock \urlprefix\url{https://link.aps.org/doi/10.1103/PhysRevB.98.121109}.

\bibitem{ZhangY18-2dm}
\bibinfo{author}{Zhang, Y.}, \bibinfo{author}{van~den Brink, J.},
  \bibinfo{author}{Felser, C.} \& \bibinfo{author}{Yan, B.}
\newblock \bibinfo{title}{Electrically tuneable nonlinear anomalous {H}all
  effect in two-dimensional transition-metal dichalcogenides {WTe}$_2$ and
  {MoTe}$_2$}.
\newblock \emph{\bibinfo{journal}{2D Materials}} \textbf{\bibinfo{volume}{5}},
  \bibinfo{pages}{044001} (\bibinfo{year}{2018}).
\newblock \urlprefix\url{http://stacks.iop.org/2053-1583/5/i=4/a=044001}.

\bibitem{Zhang18PRB}
\bibinfo{author}{Zhang, Y.}, \bibinfo{author}{Sun, Y.} \& \bibinfo{author}{Yan,
  B.}
\newblock \bibinfo{title}{Berry curvature dipole in {Weyl} semimetal materials:
  {An} ab initio study}.
\newblock \emph{\bibinfo{journal}{Phys. Rev. B}} \textbf{\bibinfo{volume}{97}},
  \bibinfo{pages}{041101} (\bibinfo{year}{2018}).
\newblock \urlprefix\url{https://link.aps.org/doi/10.1103/PhysRevB.97.041101}.

\bibitem{Du18prl}
\bibinfo{author}{Du, Z.~Z.}, \bibinfo{author}{Wang, C.~M.},
  \bibinfo{author}{Lu, H.-Z.} \& \bibinfo{author}{Xie, X.~C.}
\newblock \bibinfo{title}{Band signatures for strong nonlinear {Hall} effect in
  bilayer {WTe}$_2$}.
\newblock \emph{\bibinfo{journal}{Phys. Rev. Lett.}}
  \textbf{\bibinfo{volume}{121}}, \bibinfo{pages}{266601}
  (\bibinfo{year}{2018}).
\newblock
  \urlprefix\url{https://journals.aps.org/prl/abstract/10.1103/PhysRevLett.121.266601}.

\bibitem{Ma19nat}
\bibinfo{author}{Ma, Q.} \emph{et~al.}
\newblock \bibinfo{title}{Observation of the nonlinear {Hall} effect under
  time-reversal-symmetric conditions}.
\newblock \emph{\bibinfo{journal}{Nature}} \textbf{\bibinfo{volume}{565}},
  \bibinfo{pages}{337--342} (\bibinfo{year}{2019}).
\newblock \urlprefix\url{https://www.nature.com/articles/s41586-018-0807-6}.

\bibitem{Kang19nm}
\bibinfo{author}{Kang, K.}, \bibinfo{author}{Li, T.}, \bibinfo{author}{Sohn,
  E.}, \bibinfo{author}{Shan, J.} \& \bibinfo{author}{Mak, K.~F.}
\newblock \bibinfo{title}{Observation of the nonlinear anomalous {Hall} effect
  in few-layer {WTe}$_2$}.
\newblock \emph{\bibinfo{journal}{Nat. Mater.}} \textbf{\bibinfo{volume}{18}},
  \bibinfo{pages}{324--328} (\bibinfo{year}{2019}).
\newblock \urlprefix\url{https://www.nature.com/articles/s41563-019-0294-7}.

\bibitem{Hamamoto2017PRB}
\bibinfo{author}{Hamamoto, K.}, \bibinfo{author}{Ezawa, M.},
  \bibinfo{author}{Kim, K.~W.}, \bibinfo{author}{Morimoto, T.} \&
  \bibinfo{author}{Nagaosa, N.}
\newblock \bibinfo{title}{Nonlinear spin current generation in
  noncentrosymmetric spin-orbit coupled systems}.
\newblock \emph{\bibinfo{journal}{Phys. Rev. B}} \textbf{\bibinfo{volume}{95}},
  \bibinfo{pages}{224430} (\bibinfo{year}{2017}).
\newblock \urlprefix\url{https://link.aps.org/doi/10.1103/PhysRevB.95.224430}.

\bibitem{Araki2018SR}
\bibinfo{author}{Araki, Y.}
\newblock \bibinfo{title}{Strain-induced nonlinear spin {Hall} effect in
  topological {Dirac} semimetal}.
\newblock \emph{\bibinfo{journal}{Sci. Rep.}} \textbf{\bibinfo{volume}{8}},
  \bibinfo{pages}{1--7} (\bibinfo{year}{2018}).
\newblock \urlprefix\url{https://www.nature.com/articles/s41598-018-33655-w}.

\bibitem{Konig19prb}
\bibinfo{author}{K\"onig, E.~J.}, \bibinfo{author}{Dzero, M.},
  \bibinfo{author}{Levchenko, A.} \& \bibinfo{author}{Pesin, D.~A.}
\newblock \bibinfo{title}{Gyrotropic {H}all effect in {B}erry-curved
  materials}.
\newblock \emph{\bibinfo{journal}{Phys. Rev. B}} \textbf{\bibinfo{volume}{99}},
  \bibinfo{pages}{155404} (\bibinfo{year}{2019}).
\newblock \urlprefix\url{https://link.aps.org/doi/10.1103/PhysRevB.99.155404}.

\bibitem{Papaj19prl}
\bibinfo{author}{Papaj, M.} \& \bibinfo{author}{Fu, L.}
\newblock \bibinfo{title}{Magnus {H}all effect}.
\newblock \emph{\bibinfo{journal}{Phys. Rev. Lett.}}
  \textbf{\bibinfo{volume}{123}}, \bibinfo{pages}{216802}
  (\bibinfo{year}{2019}).
\newblock
  \urlprefix\url{https://link.aps.org/doi/10.1103/PhysRevLett.123.216802}.

\bibitem{Su19prb}
\bibinfo{author}{Yu, X.-Q.}, \bibinfo{author}{Zhu, Z.-G.},
  \bibinfo{author}{You, J.-S.}, \bibinfo{author}{Low, T.} \&
  \bibinfo{author}{Su, G.}
\newblock \bibinfo{title}{Topological nonlinear anomalous {N}ernst effect in
  strained transition metal dichalcogenides}.
\newblock \emph{\bibinfo{journal}{Phys. Rev. B}} \textbf{\bibinfo{volume}{99}},
  \bibinfo{pages}{201410} (\bibinfo{year}{2019}).
\newblock \urlprefix\url{https://link.aps.org/doi/10.1103/PhysRevB.99.201410}.

\bibitem{Zeng19prb}
\bibinfo{author}{Zeng, C.}, \bibinfo{author}{Nandy, S.},
  \bibinfo{author}{Taraphder, A.} \& \bibinfo{author}{Tewari, S.}
\newblock \bibinfo{title}{Nonlinear {N}ernst effect in bilayer {WTe}$_2$}.
\newblock \emph{\bibinfo{journal}{Phys. Rev. B}}
  \textbf{\bibinfo{volume}{100}}, \bibinfo{pages}{245102}
  (\bibinfo{year}{2019}).
\newblock \urlprefix\url{https://link.aps.org/doi/10.1103/PhysRevB.100.245102}.

\bibitem{Mandal2020PRB}
\bibinfo{author}{Mandal, D.}, \bibinfo{author}{Das, K.} \&
  \bibinfo{author}{Agarwal, A.}
\newblock \bibinfo{title}{Magnus {Nernst} and thermal {Hall} effect}.
\newblock \emph{\bibinfo{journal}{Phys. Rev. B}}
  \textbf{\bibinfo{volume}{102}}, \bibinfo{pages}{205414}
  (\bibinfo{year}{2020}).
\newblock \urlprefix\url{https://link.aps.org/doi/10.1103/PhysRevB.102.205414}.

\bibitem{Nakai2019PRB}
\bibinfo{author}{Nakai, R.} \& \bibinfo{author}{Nagaosa, N.}
\newblock \bibinfo{title}{Nonreciprocal thermal and thermoelectric transport of
  electrons in noncentrosymmetric crystals}.
\newblock \emph{\bibinfo{journal}{Phys. Rev. B}} \textbf{\bibinfo{volume}{99}},
  \bibinfo{pages}{115201} (\bibinfo{year}{2019}).
\newblock \urlprefix\url{https://link.aps.org/doi/10.1103/PhysRevB.99.115201}.

\bibitem{Zeng2020PRR}
\bibinfo{author}{Zeng, C.}, \bibinfo{author}{Nandy, S.} \&
  \bibinfo{author}{Tewari, S.}
\newblock \bibinfo{title}{Fundamental relations for anomalous thermoelectric
  transport coefficients in the nonlinear regime}.
\newblock \emph{\bibinfo{journal}{Phys. Rev. Research}}
  \textbf{\bibinfo{volume}{2}}, \bibinfo{pages}{032066} (\bibinfo{year}{2020}).
\newblock
  \urlprefix\url{https://link.aps.org/doi/10.1103/PhysRevResearch.2.032066}.

\bibitem{Nagaosa2018NCR}
\bibinfo{author}{Tokura, Y.} \& \bibinfo{author}{Nagaosa, N.}
\newblock \bibinfo{title}{Nonreciprocal responses from non-centrosymmetric
  quantum materials}.
\newblock \emph{\bibinfo{journal}{Nat. Commun.}} \textbf{\bibinfo{volume}{9}},
  \bibinfo{pages}{1--14} (\bibinfo{year}{2018}).
\newblock \urlprefix\url{https://www.nature.com/articles/s41578-020-0208-y}.

\bibitem{Sun2021NRP}
\bibinfo{author}{Du, L.} \emph{et~al.}
\newblock \bibinfo{title}{Engineering symmetry breaking in {2D} layered
  materials}.
\newblock \emph{\bibinfo{journal}{Nat. Rev. Phys.}}  (\bibinfo{year}{2021}).
\newblock \urlprefix\url{https://doi.org/10.1038/s42254-020-00276-0}.

\bibitem{Xiao10rmp}
\bibinfo{author}{Xiao, D.}, \bibinfo{author}{Chang, M.~C.} \&
  \bibinfo{author}{Niu, Q.}
\newblock \bibinfo{title}{Berry phase effects on electronic properties}.
\newblock \emph{\bibinfo{journal}{Rev. Mod. Phys.}}
  \textbf{\bibinfo{volume}{82}}, \bibinfo{pages}{1959--2007}
  (\bibinfo{year}{2010}).
\newblock \urlprefix\url{http://link.aps.org/doi/10.1103/RevModPhys.82.1959}.

\bibitem{Shvetsov19jetp}
\bibinfo{author}{Shvetsov, O.~O.}, \bibinfo{author}{Esin, V.~D.},
  \bibinfo{author}{Timonina, A.~V.}, \bibinfo{author}{Kolesnikov, N.~N.} \&
  \bibinfo{author}{Deviatov, E.~V.}
\newblock \bibinfo{title}{Nonlinear {H}all effect in three-dimensional {W}eyl
  and {D}irac semimetals}.
\newblock \emph{\bibinfo{journal}{JETP Lett.}} \textbf{\bibinfo{volume}{109}},
  \bibinfo{pages}{715--721} (\bibinfo{year}{2019}).
\newblock \urlprefix\url{https://doi.org/10.1134/S0021364019110018}.

\bibitem{Dzsaber21PNAS}
\bibinfo{author}{Dzsaber, S.} \emph{et~al.}
\newblock \bibinfo{title}{Giant spontaneous {Hall} effect in a nonmagnetic
  {Weyl{\textendash}Kondo} semimetal}.
\newblock \emph{\bibinfo{journal}{Proceedings of the National Academy of
  Sciences}} \textbf{\bibinfo{volume}{118}} (\bibinfo{year}{2021}).
\newblock \urlprefix\url{https://www.pnas.org/content/118/8/e2013386118}.

\bibitem{Qin2021CPL}
\bibinfo{author}{Qin, M.-S.} \emph{et~al.}
\newblock \bibinfo{title}{Strain tunable {Berry} curvature dipole, orbital
  magnetization and nonlinear {Hall} effect in {${\mathrm{WSe}}_{2}$}
  monolayer}.
\newblock \emph{\bibinfo{journal}{Chinese Physics Letters}}
  \textbf{\bibinfo{volume}{38}}, \bibinfo{pages}{017301}
  (\bibinfo{year}{2021}).
\newblock \urlprefix\url{https://doi.org/10.1088/0256-307x/38/1/017301}.

\bibitem{Ho2021NE}
\bibinfo{author}{Ho, S.-C.} \emph{et~al.}
\newblock \bibinfo{title}{{Hall} effects in artificially corrugated bilayer
  graphene without breaking time-reversal symmetry}.
\newblock \emph{\bibinfo{journal}{Nat. Electron.}}
  \textbf{\bibinfo{volume}{4}}, \bibinfo{pages}{116--125}
  (\bibinfo{year}{2021}).
\newblock \urlprefix\url{https://www.nature.com/articles/s41928-021-00537-5}.

\bibitem{Huang2020arXiv}
\bibinfo{author}{Huang, M.} \emph{et~al.}
\newblock \bibinfo{title}{Giant nonlinear {Hall} effect in twisted {WSe$_2$}}.
\newblock \emph{\bibinfo{journal}{arXiv preprint arXiv:2006.05615}}
  (\bibinfo{year}{2020}).
\newblock \urlprefix\url{https://arxiv.org/abs/2006.05615}.

\bibitem{Tiwari21NC}
\bibinfo{author}{Tiwari, A.} \emph{et~al.}
\newblock \bibinfo{title}{Giant c-axis nonlinear anomalous {Hall} effect in
  {T$_d$-MoTe$_2$ and WTe$_2$}}.
\newblock \emph{\bibinfo{journal}{Nature Commun.}}
  \textbf{\bibinfo{volume}{12}}, \bibinfo{pages}{1--8} (\bibinfo{year}{2021}).
\newblock \urlprefix\url{https://www.nature.com/articles/s41467-021-22343-5}.

\bibitem{Kiswandhi21arXiv}
\bibinfo{author}{Kiswandhi, A.} \& \bibinfo{author}{Osada, T.}
\newblock \bibinfo{title}{Observation of nonlinear anomalous {Hall} effect in
  organic two-dimensional {Dirac} fermion system}.
\newblock \emph{\bibinfo{journal}{arXiv:2103.00300}}  (\bibinfo{year}{2021}).
\newblock \urlprefix\url{https://arxiv.org/abs/2103.00300}.

\bibitem{He2021NC}
\bibinfo{author}{He, P.} \emph{et~al.}
\newblock \bibinfo{title}{Quantum frequency doubling in the topological
  insulator {Bi$_2$Se$_3$}}.
\newblock \emph{\bibinfo{journal}{Nat. Commun.}} \textbf{\bibinfo{volume}{12}},
  \bibinfo{pages}{1--7} (\bibinfo{year}{2021}).
\newblock \urlprefix\url{https://www.nature.com/articles/s41467-021-20983-1}.

\bibitem{Kumar2021NN}
\bibinfo{author}{Kumar, D.} \emph{et~al.}
\newblock \bibinfo{title}{Room-temperature nonlinear {Hall} effect and wireless
  radiofrequency rectification in {Weyl} semimetal {TaIrTe$_4$}}.
\newblock \emph{\bibinfo{journal}{Nat. Nanotechnol.}} \bibinfo{pages}{1--5}
  (\bibinfo{year}{2021}).
\newblock \urlprefix\url{https://www.nature.com/articles/s41565-020-00839-3}.

\bibitem{Shen17book}
\bibinfo{author}{Shen, S.-Q.}
\newblock \emph{\bibinfo{title}{Topological Insulators}}
  (\bibinfo{publisher}{Springer-Verlag}, \bibinfo{address}{Berlin Heidelberg},
  \bibinfo{year}{2017}), \bibinfo{edition}{2} edn.
\newblock
  \urlprefix\url{https://link.springer.com/book/10.1007%2F978-981-10-4606-3}.

\bibitem{Landau-ECM}
\bibinfo{author}{Landau, L.~D.} \emph{et~al.}
\newblock \emph{\bibinfo{title}{Electrodynamics of continuous media}},
  vol.~\bibinfo{volume}{8} (\bibinfo{publisher}{Elsevier,Oxford},
  \bibinfo{year}{2008}), \bibinfo{edition}{2} edn.

\bibitem{Kubo1957}
\bibinfo{author}{Kubo, R.}
\newblock \bibinfo{title}{Statistical-mechanical theory of irreversible
  processes. {I}. {General} theory and simple applications to magnetic and
  conduction problems}.
\newblock \emph{\bibinfo{journal}{J. Phys. Soc. Jpn.}}
  \textbf{\bibinfo{volume}{12}}, \bibinfo{pages}{570--586}
  (\bibinfo{year}{1957}).
\newblock \urlprefix\url{https://journals.jps.jp/doi/abs/10.1143/JPSJ.12.570}.

\bibitem{Evans93PRL}
\bibinfo{author}{Evans, D.~J.}, \bibinfo{author}{Cohen, E. G.~D.} \&
  \bibinfo{author}{Morriss, G.~P.}
\newblock \bibinfo{title}{Probability of second law violations in shearing
  steady states}.
\newblock \emph{\bibinfo{journal}{Phys. Rev. Lett.}}
  \textbf{\bibinfo{volume}{71}}, \bibinfo{pages}{2401--2404}
  (\bibinfo{year}{1993}).
\newblock \urlprefix\url{https://link.aps.org/doi/10.1103/PhysRevLett.71.2401}.

\bibitem{Evans2008}
\bibinfo{author}{J~Evans, D.} \& \bibinfo{author}{P~Morriss, G.}
\newblock \emph{\bibinfo{title}{{Statistical Mechanics of Nonequilbrium
  Liquids}}} (\bibinfo{publisher}{Cambridge University Press, Cambridge},
  \bibinfo{year}{2008}).

\bibitem{Esposito09RMP}
\bibinfo{author}{Esposito, M.}, \bibinfo{author}{Harbola, U.} \&
  \bibinfo{author}{Mukamel, S.}
\newblock \bibinfo{title}{Nonequilibrium fluctuations, fluctuation theorems,
  and counting statistics in quantum systems}.
\newblock \emph{\bibinfo{journal}{Rev. Mod. Phys.}}
  \textbf{\bibinfo{volume}{81}}, \bibinfo{pages}{1665--1702}
  (\bibinfo{year}{2009}).
\newblock \urlprefix\url{https://link.aps.org/doi/10.1103/RevModPhys.81.1665}.

\bibitem{Campisi11RMP}
\bibinfo{author}{Campisi, M.}, \bibinfo{author}{H\"anggi, P.} \&
  \bibinfo{author}{Talkner, P.}
\newblock \bibinfo{title}{{Colloquium: Quantum fluctuation relations:
  Foundations and applications}}.
\newblock \emph{\bibinfo{journal}{Rev. Mod. Phys.}}
  \textbf{\bibinfo{volume}{83}}, \bibinfo{pages}{771--791}
  (\bibinfo{year}{2011}).
\newblock \urlprefix\url{https://link.aps.org/doi/10.1103/RevModPhys.83.771}.

\bibitem{Morimoto18SR}
\bibinfo{author}{Morimoto, T.} \& \bibinfo{author}{Nagaosa, N.}
\newblock \bibinfo{title}{Nonreciprocal current from electron interactions in
  noncentrosymmetric crystals: roles of time reversal symmetry and
  dissipation}.
\newblock \emph{\bibinfo{journal}{Scientific reports}}
  \textbf{\bibinfo{volume}{8}}, \bibinfo{pages}{1--16} (\bibinfo{year}{2018}).
\newblock \urlprefix\url{https://link.aps.org/doi/10.1103/RevModPhys.83.771}.

\bibitem{Berry84}
\bibinfo{author}{Berry, M.~V.}
\newblock \bibinfo{title}{Quantal phase factors accompanying adiabatic
  changes}.
\newblock \emph{\bibinfo{journal}{Proc. R. Soc. A}}
  \textbf{\bibinfo{volume}{392}}, \bibinfo{pages}{45--57}
  (\bibinfo{year}{1984}).
\newblock
  \urlprefix\url{http://rspa.royalsocietypublishing.org/content/392/1802/45.abstract}.

\bibitem{Bohm2003Book}
\bibinfo{author}{Bohm, A.}, \bibinfo{author}{Mostafazadeh, A.},
  \bibinfo{author}{Koizumi, H.}, \bibinfo{author}{Niu, Q.} \&
  \bibinfo{author}{Zwanziger, J.}
\newblock \emph{\bibinfo{title}{{The Geometric Phase in Quantum Systems:
  Foundations, Mathematical Concepts, and Applications in Molecular and
  Condensed Matter Physics}}} (\bibinfo{publisher}{Springer-Verlag, Berlin},
  \bibinfo{year}{2003}).
\newblock
  \urlprefix\url{https://link.springer.com/book/10.1007%2F978-3-662-10333-3}.

\bibitem{Karplus1954PR}
\bibinfo{author}{Karplus, R.} \& \bibinfo{author}{Luttinger, J.~M.}
\newblock \bibinfo{title}{Hall effect in ferromagnetics}.
\newblock \emph{\bibinfo{journal}{Phys. Rev.}} \textbf{\bibinfo{volume}{95}},
  \bibinfo{pages}{1154--1160} (\bibinfo{year}{1954}).
\newblock \urlprefix\url{https://link.aps.org/doi/10.1103/PhysRev.95.1154}.

\bibitem{Thouless82prl}
\bibinfo{author}{Thouless, D.~J.}, \bibinfo{author}{Kohmoto, M.},
  \bibinfo{author}{Nightingale, M.~P.} \& \bibinfo{author}{den Nijs, M.}
\newblock \bibinfo{title}{Quantized {Hall} conductance in a two-dimensional
  periodic potential}.
\newblock \emph{\bibinfo{journal}{Phys. Rev. Lett.}}
  \textbf{\bibinfo{volume}{49}}, \bibinfo{pages}{405--408}
  (\bibinfo{year}{1982}).
\newblock \urlprefix\url{http://link.aps.org/doi/10.1103/PhysRevLett.49.405}.

\bibitem{Xu2018NP}
\bibinfo{author}{Xu, S.-Y.} \emph{et~al.}
\newblock \bibinfo{title}{Electrically switchable {Berry} curvature dipole in
  the monolayer topological insulator {WTe$_2$}}.
\newblock \emph{\bibinfo{journal}{Nat. Phys.}} \textbf{\bibinfo{volume}{14}},
  \bibinfo{pages}{900--906} (\bibinfo{year}{2018}).
\newblock \urlprefix\url{https://www.nature.com/articles/s41567-018-0189-6}.

\bibitem{Son2019PRL}
\bibinfo{author}{Son, J.}, \bibinfo{author}{Kim, K.-H.}, \bibinfo{author}{Ahn,
  Y.~H.}, \bibinfo{author}{Lee, H.-W.} \& \bibinfo{author}{Lee, J.}
\newblock \bibinfo{title}{Strain engineering of the {Berry} curvature dipole
  and valley magnetization in monolayer {${\mathrm{MoS}}_{2}$}}.
\newblock \emph{\bibinfo{journal}{Phys. Rev. Lett.}}
  \textbf{\bibinfo{volume}{123}}, \bibinfo{pages}{036806}
  (\bibinfo{year}{2019}).
\newblock
  \urlprefix\url{https://link.aps.org/doi/10.1103/PhysRevLett.123.036806}.

\bibitem{Battilomo2019PRL}
\bibinfo{author}{Battilomo, R.}, \bibinfo{author}{Scopigno, N.} \&
  \bibinfo{author}{Ortix, C.}
\newblock \bibinfo{title}{Berry curvature dipole in strained graphene: {A}
  {Fermi} surface warping effect}.
\newblock \emph{\bibinfo{journal}{Phys. Rev. Lett.}}
  \textbf{\bibinfo{volume}{123}}, \bibinfo{pages}{196403}
  (\bibinfo{year}{2019}).
\newblock
  \urlprefix\url{https://link.aps.org/doi/10.1103/PhysRevLett.123.196403}.

\bibitem{Chen2019spin}
\bibinfo{author}{Chen, C.}, \bibinfo{author}{Wang, H.}, \bibinfo{author}{Wang,
  D.} \& \bibinfo{author}{Zhang, H.}
\newblock \bibinfo{title}{Strain-engineered nonlinear {Hall} effect in {HgTe}}.
\newblock \emph{\bibinfo{journal}{Spin}} \textbf{\bibinfo{volume}{9}},
  \bibinfo{pages}{1940017} (\bibinfo{year}{2019}).
\newblock
  \urlprefix\url{https://www.worldscientific.com/doi/abs/10.1142/S2010324719400174}.

\bibitem{ZhouT20prap}
\bibinfo{author}{Zhou, B.~T.}, \bibinfo{author}{Zhang, C.-P.} \&
  \bibinfo{author}{Law, K.~T.}
\newblock \bibinfo{title}{Highly tunable nonlinear {H}all effects induced by
  spin-orbit couplings in strained polar transition-metal dichalcogenides}.
\newblock \emph{\bibinfo{journal}{Phys. Rev. Applied}}
  \textbf{\bibinfo{volume}{13}}, \bibinfo{pages}{024053}
  (\bibinfo{year}{2020}).
\newblock
  \urlprefix\url{https://link.aps.org/doi/10.1103/PhysRevApplied.13.024053}.

\bibitem{Singh20PRL}
\bibinfo{author}{Singh, S.}, \bibinfo{author}{Kim, J.}, \bibinfo{author}{Rabe,
  K.~M.} \& \bibinfo{author}{Vanderbilt, D.}
\newblock \bibinfo{title}{Engineering {W}eyl phases and nonlinear {H}all
  effects in {${\mathrm{T}}_{d}$-${\mathrm{MoTe}}_{2}$}}.
\newblock \emph{\bibinfo{journal}{Phys. Rev. Lett.}}
  \textbf{\bibinfo{volume}{125}}, \bibinfo{pages}{046402}
  (\bibinfo{year}{2020}).
\newblock
  \urlprefix\url{https://link.aps.org/doi/10.1103/PhysRevLett.125.046402}.

\bibitem{Xiao20PRApp}
\bibinfo{author}{Xiao, R.-C.}, \bibinfo{author}{Shao, D.-F.},
  \bibinfo{author}{Zhang, Z.-Q.} \& \bibinfo{author}{Jiang, H.}
\newblock \bibinfo{title}{Two-dimensional metals for piezoelectriclike devices
  based on {Berry}-curvature dipole}.
\newblock \emph{\bibinfo{journal}{Phys. Rev. Applied}}
  \textbf{\bibinfo{volume}{13}}, \bibinfo{pages}{044014}
  (\bibinfo{year}{2020}).
\newblock
  \urlprefix\url{https://link.aps.org/doi/10.1103/PhysRevApplied.13.044014}.

\bibitem{Law2020arXivWSe}
\bibinfo{author}{Hu, J.-X.}, \bibinfo{author}{Zhang, C.-P.},
  \bibinfo{author}{Xie, Y.-M.} \& \bibinfo{author}{Law, K.~T.}
\newblock \bibinfo{title}{Nonlinear {Hall} effects in strained twisted bilayer
  {WSe$_2$}}.
\newblock \emph{\bibinfo{journal}{arXiv:2004.14140}}  (\bibinfo{year}{2020}).
\newblock \urlprefix\url{https://arxiv.org/abs/2010.11086}.

\bibitem{Law2020arXivTBG}
\bibinfo{author}{Zhang, C.-P.} \emph{et~al.}
\newblock \bibinfo{title}{Giant nonlinear {Hall} effect in strained twisted
  bilayer graphene}.
\newblock \emph{\bibinfo{journal}{arXiv:2010.08333}}  (\bibinfo{year}{2020}).
\newblock \urlprefix\url{https://arxiv.org/abs/2010.08333}.

\bibitem{Pantaleon2020PRB}
\bibinfo{author}{Pantale\'on, P.~A.}, \bibinfo{author}{Low, T.} \&
  \bibinfo{author}{Guinea, F.}
\newblock \bibinfo{title}{Tunable large {Berry} dipole in strained twisted
  bilayer graphene}.
\newblock \emph{\bibinfo{journal}{Phys. Rev. B}}
  \textbf{\bibinfo{volume}{103}}, \bibinfo{pages}{205403}
  (\bibinfo{year}{2021}).
\newblock \urlprefix\url{https://link.aps.org/doi/10.1103/PhysRevB.103.205403}.

\bibitem{Weng21arXiv}
\bibinfo{author}{He, Z.} \& \bibinfo{author}{Weng, H.}
\newblock \bibinfo{title}{Giant nonlinear {Hall} effect in twisted bilayer
  {WTe$_2$}}.
\newblock \emph{\bibinfo{journal}{arXiv:2104.14288}}  (\bibinfo{year}{2021}).
\newblock \urlprefix\url{https://arxiv.org/abs/2104.14288}.

\bibitem{Isobe20sa}
\bibinfo{author}{Isobe, H.}, \bibinfo{author}{Xu, S.-Y.} \&
  \bibinfo{author}{Fu, L.}
\newblock \bibinfo{title}{High-frequency rectification via chiral {B}loch
  electrons}.
\newblock \emph{\bibinfo{journal}{Sci. Adv.}} \textbf{\bibinfo{volume}{6}},
  \bibinfo{pages}{eaay2497} (\bibinfo{year}{2020}).
\newblock
  \urlprefix\url{https://advances.sciencemag.org/content/6/13/eaay2497}.

\bibitem{Smit1955}
\bibinfo{author}{Smit, J.}
\newblock \bibinfo{title}{The spontaneous {Hall} effect in ferromagnetics {I}}.
\newblock \emph{\bibinfo{journal}{Physica}} \textbf{\bibinfo{volume}{21}},
  \bibinfo{pages}{877--887} (\bibinfo{year}{1955}).
\newblock
  \urlprefix\url{https://www.sciencedirect.com/science/article/abs/pii/S0031891455925969}.

\bibitem{Smit1958}
\bibinfo{author}{Smit, J.}
\newblock \bibinfo{title}{The spontaneous {Hall} effect in ferromagnetics
  {II}}.
\newblock \emph{\bibinfo{journal}{Physica}} \textbf{\bibinfo{volume}{24}},
  \bibinfo{pages}{39--51} (\bibinfo{year}{1958}).
\newblock
  \urlprefix\url{https://www.sciencedirect.com/science/article/abs/pii/S0031891458935419}.

\bibitem{Berger1970PRB}
\bibinfo{author}{Berger, L.}
\newblock \bibinfo{title}{Side-jump mechanism for the {Hall} effect of
  ferromagnets}.
\newblock \emph{\bibinfo{journal}{Phys. Rev. B}} \textbf{\bibinfo{volume}{2}},
  \bibinfo{pages}{4559--4566} (\bibinfo{year}{1970}).
\newblock \urlprefix\url{https://link.aps.org/doi/10.1103/PhysRevB.2.4559}.

\bibitem{Bruno2001PRB}
\bibinfo{author}{Cr\'epieux, A.} \& \bibinfo{author}{Bruno, P.}
\newblock \bibinfo{title}{Theory of the anomalous {Hall} effect from the {Kubo}
  formula and the {Dirac} equation}.
\newblock \emph{\bibinfo{journal}{Phys. Rev. B}} \textbf{\bibinfo{volume}{64}},
  \bibinfo{pages}{014416} (\bibinfo{year}{2001}).
\newblock \urlprefix\url{https://link.aps.org/doi/10.1103/PhysRevB.64.014416}.

\bibitem{Sinitsyn07jpcm}
\bibinfo{author}{Sinitsyn, N.}
\newblock \bibinfo{title}{Semiclassical theories of the anomalous {H}all
  effect}.
\newblock \emph{\bibinfo{journal}{J. Phys.: Condens. Matter}}
  \textbf{\bibinfo{volume}{20}}, \bibinfo{pages}{023201}
  (\bibinfo{year}{2008}).
\newblock
  \urlprefix\url{http://iopscience.iop.org/article/10.1088/0953-8984/20/02/023201/meta}.

\bibitem{Sinitsyn07prb}
\bibinfo{author}{Sinitsyn, N.}, \bibinfo{author}{MacDonald, A.},
  \bibinfo{author}{Jungwirth, T.}, \bibinfo{author}{Dugaev, V.} \&
  \bibinfo{author}{Sinova, J.}
\newblock \bibinfo{title}{Anomalous {H}all effect in a two-dimensional {D}irac
  band: The link between the {K}ubo-{S}treda formula and the semiclassical
  {B}oltzmann equation approach}.
\newblock \emph{\bibinfo{journal}{Phys. Rev. B}} \textbf{\bibinfo{volume}{75}},
  \bibinfo{pages}{045315} (\bibinfo{year}{2007}).
\newblock
  \urlprefix\url{https://journals.aps.org/prb/abstract/10.1103/PhysRevB.75.045315}.

\bibitem{Tian09PRL}
\bibinfo{author}{Tian, Y.}, \bibinfo{author}{Ye, L.} \& \bibinfo{author}{Jin,
  X.}
\newblock \bibinfo{title}{Proper scaling of the anomalous {Hall} effect}.
\newblock \emph{\bibinfo{journal}{Phys. Rev. Lett.}}
  \textbf{\bibinfo{volume}{103}}, \bibinfo{pages}{087206}
  (\bibinfo{year}{2009}).
\newblock
  \urlprefix\url{https://journals.aps.org/prl/abstract/10.1103/PhysRevLett.103.087206}.

\bibitem{Hou15PRL}
\bibinfo{author}{Hou, D.} \emph{et~al.}
\newblock \bibinfo{title}{Multivariable scaling for the anomalous {Hall}
  effect}.
\newblock \emph{\bibinfo{journal}{Phys. Rev. Lett.}}
  \textbf{\bibinfo{volume}{114}}, \bibinfo{pages}{217203}
  (\bibinfo{year}{2015}).
\newblock
  \urlprefix\url{https://journals.aps.org/prl/abstract/10.1103/PhysRevLett.114.217203}.

\bibitem{Yue16JPSJ}
\bibinfo{author}{Yue, D.} \& \bibinfo{author}{Jin, X.}
\newblock \bibinfo{title}{Towards a better understanding of the anomalous
  {H}all effect}.
\newblock \emph{\bibinfo{journal}{J. Phys. Soc. Jpn.}}
  \textbf{\bibinfo{volume}{86}}, \bibinfo{pages}{011006}
  (\bibinfo{year}{2016}).
\newblock
  \urlprefix\url{https://journals.jps.jp/doi/abs/10.7566/JPSJ.86.011006}.

\bibitem{Ashcroft1976}
\bibinfo{author}{Ashcroft, N.~W.} \& \bibinfo{author}{Mermin, N.~D.}
\newblock \emph{\bibinfo{title}{{Solid State Physics}}}
  (\bibinfo{publisher}{Saunders, Philadelphia}, \bibinfo{year}{1976}).

\bibitem{Du19nc}
\bibinfo{author}{Du, Z.~Z.}, \bibinfo{author}{Wang, C.~M.},
  \bibinfo{author}{Li, S.}, \bibinfo{author}{Lu, H.-Z.} \&
  \bibinfo{author}{Xie, X.~C.}
\newblock \bibinfo{title}{Disorder-induced nonlinear {Hall} effect with
  time-reversal symmetry}.
\newblock \emph{\bibinfo{journal}{Nat. Commun.}} \textbf{\bibinfo{volume}{10}},
  \bibinfo{pages}{3047} (\bibinfo{year}{2019}).
\newblock \urlprefix\url{https://www.nature.com/articles/s41467-019-10941-3}.

\bibitem{Pancharatnam1956}
\bibinfo{author}{Pancharatnam, S.}
\newblock \bibinfo{title}{Generalized theory of interference and its
  applications}.
\newblock In \emph{\bibinfo{booktitle}{Proceedings of the Indian Academy of
  Sciences-Section A}}, vol.~\bibinfo{volume}{44}, \bibinfo{pages}{398--417}
  (\bibinfo{organization}{Springer}, \bibinfo{year}{1956}).
\newblock \urlprefix\url{https://link.springer.com/article/10.1007/BF03046095}.

\bibitem{Sinitsyn2006PRB}
\bibinfo{author}{Sinitsyn, N.~A.}, \bibinfo{author}{Niu, Q.} \&
  \bibinfo{author}{MacDonald, A.~H.}
\newblock \bibinfo{title}{Coordinate shift in the semiclassical {Boltzmann}
  equation and the anomalous {Hall} effect}.
\newblock \emph{\bibinfo{journal}{Phys. Rev. B}} \textbf{\bibinfo{volume}{73}},
  \bibinfo{pages}{075318} (\bibinfo{year}{2006}).
\newblock \urlprefix\url{https://link.aps.org/doi/10.1103/PhysRevB.73.075318}.

\bibitem{Resta21arXiv}
\bibinfo{author}{Resta, R.}
\newblock \bibinfo{title}{Linear and nonlinear {Hall} conductivity in presence
  of interaction and disorder}.
\newblock \emph{\bibinfo{journal}{arXiv:2101.10949}}  (\bibinfo{year}{2021}).
\newblock \urlprefix\url{https://arxiv.org/abs/2101.10949}.

\bibitem{Nandy19prb}
\bibinfo{author}{Nandy, S.} \& \bibinfo{author}{Sodemann, I.}
\newblock \bibinfo{title}{Symmetry and quantum kinetics of the nonlinear {H}all
  effect}.
\newblock \emph{\bibinfo{journal}{Phys. Rev. B}}
  \textbf{\bibinfo{volume}{100}}, \bibinfo{pages}{195117}
  (\bibinfo{year}{2019}).
\newblock \urlprefix\url{https://link.aps.org/doi/10.1103/PhysRevB.100.195117}.

\bibitem{Xiao19prb}
\bibinfo{author}{Xiao, C.}, \bibinfo{author}{Du, Z.~Z.} \&
  \bibinfo{author}{Niu, Q.}
\newblock \bibinfo{title}{Theory of nonlinear {H}all effects: Modified
  semiclassics from quantum kinetics}.
\newblock \emph{\bibinfo{journal}{Phys. Rev. B}}
  \textbf{\bibinfo{volume}{100}}, \bibinfo{pages}{165422}
  (\bibinfo{year}{2019}).
\newblock \urlprefix\url{https://link.aps.org/doi/10.1103/PhysRevB.100.165422}.

\bibitem{Du2020arXiv}
\bibinfo{author}{Du, Z.~Z.}, \bibinfo{author}{Wang, C.~M.},
  \bibinfo{author}{Sun, H.-P.}, \bibinfo{author}{Lu, H.-Z.} \&
  \bibinfo{author}{Xie, X.~C.}
\newblock \bibinfo{title}{Quantum theory of the nonlinear {Hall} effect}.
\newblock \emph{\bibinfo{journal}{arXiv preprint arXiv:2004.09742}}
  (\bibinfo{year}{2020}).
\newblock \urlprefix\url{https://arxiv.org/abs/2004.09742}.

\bibitem{Gao2020PRB}
\bibinfo{author}{Gao, Y.}, \bibinfo{author}{Zhang, F.} \&
  \bibinfo{author}{Zhang, W.}
\newblock \bibinfo{title}{Second-order nonlinear {Hall} effect in {Weyl}
  semimetals}.
\newblock \emph{\bibinfo{journal}{Phys. Rev. B}}
  \textbf{\bibinfo{volume}{102}}, \bibinfo{pages}{245116}
  (\bibinfo{year}{2020}).
\newblock \urlprefix\url{https://link.aps.org/doi/10.1103/PhysRevB.102.245116}.

\bibitem{Levchenko2021arXiv}
\bibinfo{author}{K{\"o}nig, E.~J.} \& \bibinfo{author}{Levchenko, A.}
\newblock \bibinfo{title}{Quantum kinetics of anomalous and nonlinear {Hall}
  effects in topological semimetals}.
\newblock \emph{\bibinfo{journal}{arXiv:2102.05675}}  (\bibinfo{year}{2021}).
\newblock \urlprefix\url{https://arxiv.org/abs/2102.05675}.

\bibitem{Culcer17PRB}
\bibinfo{author}{Culcer, D.}, \bibinfo{author}{Sekine, A.} \&
  \bibinfo{author}{MacDonald, A.~H.}
\newblock \bibinfo{title}{Interband coherence response to electric fields in
  crystals: Berry-phase contributions and disorder effects}.
\newblock \emph{\bibinfo{journal}{Phys. Rev. B}} \textbf{\bibinfo{volume}{96}},
  \bibinfo{pages}{035106} (\bibinfo{year}{2017}).
\newblock \urlprefix\url{https://link.aps.org/doi/10.1103/PhysRevB.96.035106}.

\bibitem{Zhao16np}
\bibinfo{author}{Zhao, L.} \emph{et~al.}
\newblock \bibinfo{title}{Evidence of an odd-parity hidden order in a
  spin--orbit coupled correlated iridate}.
\newblock \emph{\bibinfo{journal}{Nat. Phys.}} \textbf{\bibinfo{volume}{12}},
  \bibinfo{pages}{32--36} (\bibinfo{year}{2016}).
\newblock \urlprefix\url{https://www.nature.com/articles/nphys3517}.

\bibitem{Zhao17np}
\bibinfo{author}{Zhao, L.} \emph{et~al.}
\newblock \bibinfo{title}{A global inversion-symmetry-broken phase inside the
  pseudogap region of {YBa$_2$Cu$_3$O$_y$}}.
\newblock \emph{\bibinfo{journal}{Nat. Phys.}} \textbf{\bibinfo{volume}{13}},
  \bibinfo{pages}{250--254} (\bibinfo{year}{2017}).
\newblock \urlprefix\url{https://www.nature.com/articles/nphys3962}.

\bibitem{Xiao20PRB}
\bibinfo{author}{Xiao, R.-C.}, \bibinfo{author}{Shao, D.-F.},
  \bibinfo{author}{Huang, W.} \& \bibinfo{author}{Jiang, H.}
\newblock \bibinfo{title}{Electrical detection of ferroelectriclike metals
  through the nonlinear {Hall} effect}.
\newblock \emph{\bibinfo{journal}{Phys. Rev. B}}
  \textbf{\bibinfo{volume}{102}}, \bibinfo{pages}{024109}
  (\bibinfo{year}{2020}).
\newblock \urlprefix\url{https://link.aps.org/doi/10.1103/PhysRevB.102.024109}.

\bibitem{Habib20prr}
\bibinfo{author}{Rostami, H.} \& \bibinfo{author}{Juri\ifmmode \check{c}\else
  \v{c}\fi{}i\ifmmode~\acute{c}\else \'{c}\fi{}, V.}
\newblock \bibinfo{title}{Probing quantum criticality using nonlinear {H}all
  effect in a metallic dirac system}.
\newblock \emph{\bibinfo{journal}{Phys. Rev. Research}}
  \textbf{\bibinfo{volume}{2}}, \bibinfo{pages}{013069} (\bibinfo{year}{2020}).
\newblock
  \urlprefix\url{https://link.aps.org/doi/10.1103/PhysRevResearch.2.013069}.

\bibitem{Shao20prl}
\bibinfo{author}{Shao, D.-F.}, \bibinfo{author}{Zhang, S.-H.},
  \bibinfo{author}{Gurung, G.}, \bibinfo{author}{Yang, W.} \&
  \bibinfo{author}{Tsymbal, E.~Y.}
\newblock \bibinfo{title}{Nonlinear anomalous {H}all effect for {N}\'eel vector
  detection}.
\newblock \emph{\bibinfo{journal}{Phys. Rev. Lett.}}
  \textbf{\bibinfo{volume}{124}}, \bibinfo{pages}{067203}
  (\bibinfo{year}{2020}).
\newblock
  \urlprefix\url{https://link.aps.org/doi/10.1103/PhysRevLett.124.067203}.

\bibitem{Xiao2020NP}
\bibinfo{author}{Xiao, J.} \emph{et~al.}
\newblock \bibinfo{title}{Berry curvature memory through electrically driven
  stacking transitions}.
\newblock \emph{\bibinfo{journal}{Nat. Phys.}} \textbf{\bibinfo{volume}{16}},
  \bibinfo{pages}{1028--1034} (\bibinfo{year}{2020}).
\newblock \urlprefix\url{https://www.nature.com/articles/s41567-020-0947-0}.

\bibitem{Fu2021arXiv}
\bibinfo{author}{Zhang, Y.} \& \bibinfo{author}{Fu, L.}
\newblock \bibinfo{title}{Terahertz detection based on nonlinear {Hall} effect
  without magnetic field}.
\newblock \emph{\bibinfo{journal}{arXiv:2101.05842}}  (\bibinfo{year}{2021}).
\newblock \urlprefix\url{https://arxiv.org/abs/2101.05842}.

\bibitem{Kim17PRB}
\bibinfo{author}{Kim, K.~W.}, \bibinfo{author}{Morimoto, T.} \&
  \bibinfo{author}{Nagaosa, N.}
\newblock \bibinfo{title}{Shift charge and spin photocurrents in {Dirac}
  surface states of topological insulator}.
\newblock \emph{\bibinfo{journal}{Phys. Rev. B}} \textbf{\bibinfo{volume}{95}},
  \bibinfo{pages}{035134} (\bibinfo{year}{2017}).
\newblock \urlprefix\url{https://link.aps.org/doi/10.1103/PhysRevB.95.035134}.

\bibitem{Bhalla20PRL}
\bibinfo{author}{Bhalla, P.}, \bibinfo{author}{MacDonald, A.~H.} \&
  \bibinfo{author}{Culcer, D.}
\newblock \bibinfo{title}{Resonant photovoltaic effect in doped magnetic
  semiconductors}.
\newblock \emph{\bibinfo{journal}{Phys. Rev. Lett.}}
  \textbf{\bibinfo{volume}{124}}, \bibinfo{pages}{087402}
  (\bibinfo{year}{2020}).
\newblock
  \urlprefix\url{https://link.aps.org/doi/10.1103/PhysRevLett.124.087402}.

\bibitem{Toshio2020PRR}
\bibinfo{author}{Toshio, R.}, \bibinfo{author}{Takasan, K.} \&
  \bibinfo{author}{Kawakami, N.}
\newblock \bibinfo{title}{Anomalous hydrodynamic transport in interacting
  noncentrosymmetric metals}.
\newblock \emph{\bibinfo{journal}{Phys. Rev. Research}}
  \textbf{\bibinfo{volume}{2}}, \bibinfo{pages}{032021} (\bibinfo{year}{2020}).
\newblock
  \urlprefix\url{https://link.aps.org/doi/10.1103/PhysRevResearch.2.032021}.

\bibitem{Law2020arXivHigher}
\bibinfo{author}{Zhang, C.-P.}, \bibinfo{author}{Gao, X.-J.},
  \bibinfo{author}{Xie, Y.-M.}, \bibinfo{author}{Po, H.~C.} \&
  \bibinfo{author}{Law, K.~T.}
\newblock \bibinfo{title}{Higher-order nonlinear anomalous {Hall} effects
  induced by {Berry} curvature multipoles}.
\newblock \emph{\bibinfo{journal}{arXiv:2012.15628}}  (\bibinfo{year}{2020}).
\newblock \urlprefix\url{https://arxiv.org/abs/2012.15628}.

\bibitem{He2018NP}
\bibinfo{author}{He, P.} \emph{et~al.}
\newblock \bibinfo{title}{Bilinear magnetoelectric resistance as a probe of
  three-dimensional spin texture in topological surface states}.
\newblock \emph{\bibinfo{journal}{Nat. Phys.}} \textbf{\bibinfo{volume}{14}},
  \bibinfo{pages}{495--499} (\bibinfo{year}{2018}).
\newblock \urlprefix\url{https://www.nature.com/articles/s41567-017-0039-y}.

\bibitem{He2018PRL}
\bibinfo{author}{He, P.} \emph{et~al.}
\newblock \bibinfo{title}{Observation of out-of-plane spin texture in a
  {${\mathrm{SrTiO}}_{3}(111)$} two-dimensional electron gas}.
\newblock \emph{\bibinfo{journal}{Phys. Rev. Lett.}}
  \textbf{\bibinfo{volume}{120}}, \bibinfo{pages}{266802}
  (\bibinfo{year}{2018}).
\newblock
  \urlprefix\url{https://link.aps.org/doi/10.1103/PhysRevLett.120.266802}.

\bibitem{He2019NC}
\bibinfo{author}{He, P.} \emph{et~al.}
\newblock \bibinfo{title}{Nonlinear magnetotransport shaped by fermi surface
  topology and convexity}.
\newblock \emph{\bibinfo{journal}{Nat. Commun.}} \textbf{\bibinfo{volume}{10}},
  \bibinfo{pages}{1--7} (\bibinfo{year}{2019}).
\newblock \urlprefix\url{https://www.nature.com/articles/s41467-019-09208-8}.

\bibitem{He2019PRL}
\bibinfo{author}{He, P.} \emph{et~al.}
\newblock \bibinfo{title}{Nonlinear planar {Hall} effect}.
\newblock \emph{\bibinfo{journal}{Phys. Rev. Lett.}}
  \textbf{\bibinfo{volume}{123}}, \bibinfo{pages}{016801}
  (\bibinfo{year}{2019}).
\newblock
  \urlprefix\url{https://link.aps.org/doi/10.1103/PhysRevLett.123.016801}.

\bibitem{Zhang2018Spin}
\bibinfo{author}{Zhang, S. S.-L.} \& \bibinfo{author}{Vignale, G.}
\newblock \bibinfo{title}{Theory of bilinear magneto-electric resistance from
  topological-insulator surface states}.
\newblock \emph{\bibinfo{journal}{Spintronics XI}}
  \textbf{\bibinfo{volume}{10732}}, \bibinfo{pages}{1073215}
  (\bibinfo{year}{2018}).
\newblock \urlprefix\url{https://doi.org/10.1117/12.2323126}.

\bibitem{Fert2020PRL}
\bibinfo{author}{Dyrda\l{}, A.}, \bibinfo{author}{Barna\ifmmode~\acute{s}\else
  \'{s}\fi{}, J.} \& \bibinfo{author}{Fert, A.}
\newblock \bibinfo{title}{Spin-momentum-locking inhomogeneities as a source of
  bilinear magnetoresistance in topological insulators}.
\newblock \emph{\bibinfo{journal}{Phys. Rev. Lett.}}
  \textbf{\bibinfo{volume}{124}}, \bibinfo{pages}{046802}
  (\bibinfo{year}{2020}).
\newblock
  \urlprefix\url{https://link.aps.org/doi/10.1103/PhysRevLett.124.046802}.

\bibitem{Zyuzin2018PRB}
\bibinfo{author}{Zyuzin, A.~A.}, \bibinfo{author}{Silaev, M.} \&
  \bibinfo{author}{Zyuzin, V.~A.}
\newblock \bibinfo{title}{Nonlinear chiral transport in {Dirac} semimetals}.
\newblock \emph{\bibinfo{journal}{Phys. Rev. B}} \textbf{\bibinfo{volume}{98}},
  \bibinfo{pages}{205149} (\bibinfo{year}{2018}).
\newblock \urlprefix\url{https://link.aps.org/doi/10.1103/PhysRevB.98.205149}.

\bibitem{zeng2020arXivChiral}
\bibinfo{author}{Zeng, C.}, \bibinfo{author}{Nandy, S.} \&
  \bibinfo{author}{Tewari, S.}
\newblock \bibinfo{title}{Chiral anomaly induced nonlinear {Nernst} and thermal
  {Hall} effects in {Weyl} semimetals}.
\newblock \emph{\bibinfo{journal}{arXiv:2012.11590}}  (\bibinfo{year}{2020}).
\newblock \urlprefix\url{https://arxiv.org/abs/2012.11590}.

\bibitem{Zhang2021PRB}
\bibinfo{author}{Li, R.-H.}, \bibinfo{author}{Heinonen, O.~G.},
  \bibinfo{author}{Burkov, A.~A.} \& \bibinfo{author}{Zhang, S. S.-L.}
\newblock \bibinfo{title}{Nonlinear {Hall} effect in {Weyl} semimetals induced
  by chiral anomaly}.
\newblock \emph{\bibinfo{journal}{Phys. Rev. B}}
  \textbf{\bibinfo{volume}{103}}, \bibinfo{pages}{045105}
  (\bibinfo{year}{2021}).
\newblock \urlprefix\url{https://link.aps.org/doi/10.1103/PhysRevB.103.045105}.

\bibitem{Deviatov2020JETP}
\bibinfo{author}{Esin, V.~D.}, \bibinfo{author}{Timonina, A.~V.},
  \bibinfo{author}{Kolesnikov, N.~N.} \& \bibinfo{author}{Deviatov, E.~V.}
\newblock \bibinfo{title}{Second-harmonic voltage response for the magnetic
  {Weyl} semimetal {Co$_3$Sn$_2$S$_2$}}.
\newblock \emph{\bibinfo{journal}{JETP Letters}}
  \textbf{\bibinfo{volume}{111}}, \bibinfo{pages}{685--689}
  (\bibinfo{year}{2020}).
\newblock
  \urlprefix\url{https://link.springer.com/article/10.1134/S0021364020120024}.

\bibitem{Watanabe2021PRR}
\bibinfo{author}{Watanabe, H.} \& \bibinfo{author}{Yanase, Y.}
\newblock \bibinfo{title}{Nonlinear electric transport in odd-parity magnetic
  multipole systems: {Application to Mn-based compounds}}.
\newblock \emph{\bibinfo{journal}{Phys. Rev. Research}}
  \textbf{\bibinfo{volume}{2}}, \bibinfo{pages}{043081} (\bibinfo{year}{2020}).
\newblock
  \urlprefix\url{https://link.aps.org/doi/10.1103/PhysRevResearch.2.043081}.

\bibitem{Boyd1992Book}
\bibinfo{author}{Boyd, R.~W.}
\newblock \emph{\bibinfo{title}{{Nonlinear Optics}}}
  (\bibinfo{publisher}{Academic, San Diego}, \bibinfo{year}{1992}).

\bibitem{Flensberg95PRB}
\bibinfo{author}{Flensberg, K.}, \bibinfo{author}{Hu, B. Y.-K.},
  \bibinfo{author}{Jauho, A.-P.} \& \bibinfo{author}{Kinaret, J.~M.}
\newblock \bibinfo{title}{Linear-response theory of {C}oulomb drag in coupled
  electron systems}.
\newblock \emph{\bibinfo{journal}{Phys. Rev. B}} \textbf{\bibinfo{volume}{52}},
  \bibinfo{pages}{14761--14774} (\bibinfo{year}{1995}).
\newblock \urlprefix\url{https://link.aps.org/doi/10.1103/PhysRevB.52.14761}.

\bibitem{Kamenev95PRB}
\bibinfo{author}{Kamenev, A.} \& \bibinfo{author}{Oreg, Y.}
\newblock \bibinfo{title}{Coulomb drag in normal metals and superconductors:
  {D}iagrammatic approach}.
\newblock \emph{\bibinfo{journal}{Phys. Rev. B}} \textbf{\bibinfo{volume}{52}},
  \bibinfo{pages}{7516--7527} (\bibinfo{year}{1995}).
\newblock \urlprefix\url{https://link.aps.org/doi/10.1103/PhysRevB.52.7516}.

\end{thebibliography}

\end{document}